\documentclass[12pt]{article}

\usepackage{scicite}
\usepackage{graphicx}
\usepackage{times}
\usepackage{url}
\usepackage[justification=raggedright, singlelinecheck=false]{caption}

\newenvironment{sciabstract}{%
\begin{quote} \bf}
{\end{quote}}

\title{Increasing trend of scientists to switch between topics}

\author
{An Zeng$^1$, Zhesi Shen$^2$, Jianlin Zhou$^1$, Ying Fan$^1$, Zengru Di$^1$,\\ Yougui Wang$^{1,\ast}$, H. Eugene Stanley,$^{3,\ast}$ and Shlomo Havlin$^{4,\ast}$\\
\\
\normalsize{$^{1}$School of Systems Science, Beijing Normal University, Beijing 100875, China}\\
\normalsize{$^{2}$National Science Library, Chinese Academy of Sciences, Beijing 100190, China}\\
\normalsize{$^{2}$Center for Polymer Studies and Department of Physics, Boston University, Boston, MA 02215}\\
\normalsize{$^{2}$Department of Physics, Bar-Ilan University, Ramat-Gan 52900, Israel}\\
\\
\normalsize{$^\ast$To whom correspondence should be addressed;}\\
\normalsize{E-mail: ygwang@bnu.edu.cn (Y.W.), hes@bu.edu (H.E.S.), havlin@ophir.ph.biu.ac.il (S.H.)}
}

\date{}

\begin{document}


\baselineskip24pt


\maketitle


\begin{sciabstract}
We analyze the publication records of individual scientists, aiming to quantify the topic switching dynamics of scientists and its influence. For each scientist, the relations among her publications are characterized via shared references. We find that the co-citing network of the papers of a scientist exhibits a clear community structure where each major community represents a research topic. Our analysis suggests that scientists tend to have a narrow distribution of the number of topics. However, researchers nowadays switch more frequently between topics than those in the early days. We also find that high switching probability in early career ($<12y$) is associated with low overall productivity, while it is correlated with high overall productivity in latter career. Interestingly, the average citation per paper, however, is in all career stages negatively correlated with the switching probability. We propose a model with exploitation and exploration mechanisms that can explain the main observed features.
\end{sciabstract}

\section*{Introduction}

Uncovering the mechanisms governing research activities of individual scientists and their evolution with time is critical for understanding and managing a wide range of issues in science, from training of scientists to collective discovery of new knowledge~\cite{the2017zeng,standing2017qi,standing2017amjad,choosing2015rzhetsky,quantifying2016domenico}. The digital publishing era has led to a revolution in science embodied in big data that captures major activities in research. This creates an unprecedented opportunity to explore the dynamical patterns of scientific production and reward using state-of-the-art mathematical and computational tools~\cite{data2017clauset,science2018fortunato,inheritance2014kuhn}. Apart from the early works aiming at evaluating scientific impact with scientists' citations~\cite{networks1965price}, $h$-index~\cite{an2005hirsch} and related variants~\cite{review2016waltman}, there is a recent wave of studies focusing on quantifying and modeling the evolution of research creativity throughout scientists' careers~\cite{persistence2012petersen,quantifying2016sinatra,hot2018liu,age2011jones,reputation2014petersen,quantifying2015petersen,career2014deville,multiscale2018petersen}. Scientists' cumulative production has been shown to exhibit persistent growth with time~\cite{persistence2012petersen}, which is associated with the well-known Matthew effect~\cite{the1968merton}. By associating each publication with its citation, it has been revealed that the most influential work of a scientist appears randomly within the sequence of her publications~\cite{quantifying2016sinatra}. A follow-up work investigated the timing of top-$n$ most influential papers of an individual researcher, revealing that scientists' career may involve a hot streak period during which an individual's performance is substantially higher than her typical performance~\cite{hot2018liu}. Other issues such as the evolution of scientists' creativity~\cite{age2011jones}, reputation~\cite{reputation2014petersen}, social ties~\cite{quantifying2015petersen} and mobility~\cite{career2014deville,multiscale2018petersen} over their careers have also been investigated.

A fundamental driving force of scientific research is the evolution of scientists' research interest~\cite{quantify2016domenico}, which is reflected in the switching of scientists between different research topics over time. Sociologists of science have made persistent effort in qualitative understanding the principles governing the topic selection of scientists, and pointed out that it may result from a trade-off between conservative production and risky innovation~\cite{the1975bourdieu}. There are also rich illustrative models proposed by sociologists to categorize the research strategies adopted by scientists~\cite{tradition2016foster}. With the increasing availability of scientific publication data, the issue of topic selection started to be analyzed quantitatively in recent years. It has been pointed out that the research interest of individual physicists could shift significantly from the beginning to the end of the career, with the distance between interests being measured based on field classification codes in physics~\cite{quantifying2017jia}. However, the variation of topic switching during the individual career has not been studied so far. Here we ask: How to identify the topics that an individual scientist is involved? How frequent a scientist switches between different research topics? Does more frequent switching of scientists between topics help their impact? Does the topic switching behavior of scientists change during the past century?

To address these questions, we construct a network for each scientist characterizing the relations between her papers. The structure of this network will immediately reveal how an individual scientist's research interests are embodied. This framework allows us, applying community analysis, to specify the various research interests and accordingly investigate the detailed dynamics of the research interest shifting of a scientist, as well as the switching tendency evolution during the last century and its relation to research impact. Our analysis suggests that scientists tend to have a narrow distribution of the number of major topics during their life time. We find that the typical number of major topics during last century stays almost unchanged. However, researchers in the early days tend to work in a topic for a longer time before switching to another topic, while nowadays they tend to work on multiple topics simultaneously. Interestingly, we find that more frequent switching between topics in the early career ($<12y$) is related to lower research performance, i.e., both the overall productivity and mean citation are lower. In marked contrast, more frequent switching in the latter career is associated with higher overall productivity but with lower mean citations. We propose a model reproducing the main observed empirical patterns. Our framework, although applied here to physicists and computer scientists, is general and not restricted to availability of field classification codes, so it can be applied to analyzing scientists from any discipline.

\section*{Results}
In this paper, we analyze the scientific publication data of the American Physical Society (APS) journals. Disambiguated author name data provided in~\cite{quantifying2016sinatra} is used to assign each paper to its authors, which results in the publication records of 236,884 distinct scientists (for basic statistics of this data see Fig. S1 of Supplementary Materials (SM)). In order to investigate how the papers of an individual scientist are related, we construct for each scientist a co-citing network (CCN) in which each node is a paper authored by this scientist and two papers have a link if they share at least one reference. This approach of constructing links between nodes (papers) based on their common neighbors is called bibliographic coupling in Scientometrics~\cite{bibliographic1963kessler,co2010boyack} and has also been widely used in the analysis of various other real systems such as international trading systems~\cite{the2007hidalgo} and online social systems~\cite{bipartite2007zhou}. The communities of each co-citing network of a scientist are identified with the fast unfolding algorithm which detects communities by maximizing the modularity function~\cite{fast2008blondel}. Typically, a network contains several large-size communities as well as some small clusters and isolated nodes. The major communities represent the main research topics of this scientist. As the network size needs to be large enough to ensure meaningful community detection results, we consider in this study all scientists that have published at least 50 papers in the APS journals (3,420 scientists in total, for the distribution of their career start years see Fig. S2). Results for scientists with fewer papers (at least 20 papers, 15,373 scientists) are similar and are reported in Figs. S10 and S11 of the Supplementary Materials (SM). In addition, we have studied the communities detected in the weighted co-citing network where links are weighted according to the number of shared references. The community structure is not significantly altered when considering the link weights (see Fig. S3), as large weights tend to locate on the links within communities. Our community analysis has also been examined based on a modified modularity function with higher resolution parameter (see Figs. S12 and S13 in SM) and on another data set from computer science (see Figs. S15 and S16 in SM) and for all tests, the main conclusions have been found to be similar.

Illustration of the co-citing network of a typical highly-cited scientist is given in Fig. 1. The community connectivity matrix in Fig. 1c shows that nodes within each community are well connected, yet nodes between communities are much less connected. The time series presented in Fig. 1d describes the growth history of the network and reveals how this scientist moves from one research topic to another during his career. In the time series, each point is a paper and different colors represent different communities in the co-citing network. The height of the point is the number of links (i.e. degree) that the paper has in the network. The analyses in our study are mainly based on the co-citing networks and time series of scientists.

We first focus on the structural properties of the co-citing networks (CCNs). For each scientist's CCN, we calculate the size of its giant component (GC) and study its correlation with the network size, as shown in the scatter plot presented in Fig. 2a. It is seen that most of the points are located close to the diagonal line, indicating that CCNs are in general well connected and have relatively large GCs (see Fig. S4 in SM for the results with the network including also the co-cited relations between papers). This is also seen in the inset of Fig. 2a where a significant right-skewed distribution (close to 1) of the relative size of GC is observed. Fig. 1c suggests that a CCN has a community structure. As a statistical support for this phenomenon, we plot in Fig. 2b the maximized modularity, $Q_{real}$, in real CCNs and the maximized modularity, $Q_{rand}$, in their degree-preserved reshuffled counterparts. All points are located under the diagonal line, indicating that the community structure in real CCN is truly significant.

Given that papers tend to cluster into communities in CCN, one interesting question is what is the typical number of communities that a scientist has. We show in Fig. 2c the distribution of the number of communities for all scientists. The number of communities is seemingly broadly distributed. However, as CCNs may consist of isolated nodes or very small clusters, we use a threshold to eliminate communities that are too small to be regarded as a research field of a researcher. After filtering, the distributions of the number of communities that a scientist has become very narrow, with the peaks around 4 and 3 if communities with only sizes larger than 2 and 5 are considered respectively. In the following analysis, we define major communities as such of more than two nodes. To better understand the community size in CCNs, we show in Fig. 2d the fraction of papers in each community sorted by size in descending order. The strong decay of the curve indicates that several major communities comprise most of the nodes. A further investigation of the inverse cumulative probability of fraction of nodes in the several largest communities indicates that for half of the scientists, the three largest communities include over 70\% of their papers, as seen in Fig. 2e.

In each CCN, a major community contains papers that are topologically close to each other. In order to validate whether the papers in a community are indeed in similar research topics~\cite{centuary2015sinatra,document2018trujillo}, we analyze the PACS code (a field classification code in physics) of the papers belonging to the same community. We show in Fig. 2f the Gini coefficient~\cite{variability1912gini} of the distribution of PACS codes in different communities. A larger Gini coefficient corresponds to a more heterogeneous distribution of the PACS codes in a community. The real data is compared with a random counterpart where the PACS codes are reshuffled among each individual scientist's papers while the community structure is preserved. We show in Fig. 2f the mean Gini coefficient in each community sorted by size in descending order. We find that the mean Gini coefficient in real data is higher than that in the random counterpart, with a p-value smaller than 0.01 in the Kolmogorov-Smirnov test of the corresponding Gini coefficient distributions. Thus, our results suggest that papers in a community tend to share the same PACS codes and the detected communities reflect distinct research fields of a scientist.

Once the detected communities are marked in the time series (Fig. 1d), the dynamics of scientists' interest across different research topics can be investigated. To this end, we first show in Fig. 3a, the mean number of yearly involved major communities for each scientist. It can be seen that scientists tend to be involved in small number of communities during their early career. Then the number of yearly involved communities increases until it peaks around the $20_{th}$ year of the career, and gradually decreases after that. However, when a scientist publishes more papers in a year, she might have a higher number of yearly involved communities purely by chance. To remove the effect of number of yearly published papers (see Fig. S5 in SM), we propose another metric called switching probability which computes the probability of a scientist to switch from one major community to another major community between two adjacent publications. Fig. 3b shows the evolution of the mean switching probability in different career years. The peak of switching probability is also around the $20_{th}$ career year, indicating that scientists tend to switch less during their early career while switch more in the later stage of their career, which is consistent with the trend observed with the yearly involved communities.

We further ask, does increasing switching helps research performance or not? To this end, we investigate the correlation between the switching rate and research performance. Here, we measure the research performance of a scientist using two almost uncorrelated metrics (see Fig. S8), i.e., number of published papers and mean citation per paper. Consistent with ref.~\cite{quantifying2016sinatra}, we only consider the number of citations 10 years after a paper is published, i.e. $c_{10}$. We first compare in Fig. 3c, the overall switching probability with the switching probability of the 10\% most productive scientists in different career years. We find surprisingly two opposite behaviors. In the early career stage ($<12y$) high overall productivity is associated with low switching probability yet in later career stage high productivity is associated with higher switching probability. In addition, we compare in Fig. 3d, the overall switching probability with the switching probability of the 10\% scientists who has the highest mean citation per paper. The figure shows that \emph{high} average citation per paper in all career periods is associated with \emph{low} switching probability. This interesting finding might be due to the fact that higher switching probability reduces the impression of leadership in a specific field, yielding less citations. This result is also highly supported by an additional test where the switching probability is found to be negatively correlated with mean citation per paper, especially for productive scientists (see Fig. S9 in SM). To examine the significance of these findings, we carry out the Kolmogorov-Smirnov test of the switching probability distribution in each career year. The small p-value shown in the insets of Figs. 3c and 3d (mostly $p<0.05$) suggests that the overall (total population) switching probability indeed follows a \emph{distinct} distribution from each of the two sub-groups of scientists (i.e. 10\% most productive and 10\% most highly cited per paper) in each career year. We additionally calculate the Pearson correlation between scientists' switching probability in different career years and their overall productivity, as well as the Pearson correlation between scientists' switching probability in different career years and the mean citation per paper. The correlations presented in Fig. S6a and S6b also highly support the findings revealed in Fig. 3c and 3d.

Next we study how the structural and dynamical properties of CCNs evolve as the development of science in the last 100 years. As our data ends in 2010, the careers of some scientists are not completed. We thus have to fix the career length of the scientists from different years in order to ensure a fair comparison between their CCNs. Specifically, we only consider scientists' first $y$ career years and remove (i) all the scientists who did not yet reach $y$ years career and (ii) those who published less than 30 papers in their first $y$ career years. In our analysis, we present results of $y=10, 20, 30$. We first select the scientists who started their careers in a certain year and average the number of major communities that these scientists have been involved in their careers. We show in Fig. 4a the mean number of communities for the scientists who started their career in different years. The results indicate that as science evolves, the number of major communities of individual scientists stays almost unchanged. The evolution of other structural properties of CCNs is presented in Fig. S7. We further calculate the mean switching probability of each scientist over her career and accordingly compute the mean switching probability per year by averaging the switching probability of all scientists who started their career in this year. We show in Fig. 4b the average switching probability of scientists who started their career in different years. The results surprisingly indicate that although the number of communities is stable over years, scientists tend to increase switching between communities, i.e., topics, during last century. More specifically, scientists in the earlier days tend to work in a topic for a longer period before switching to another topic. On the contrary, scientists nowadays tend to work on multiple topics almost simultaneously, resulting in more frequent switching between communities almost in each pair of adjacent publications. We then test the significance of our observed trends by directly studying the distributions of number of communities and the switching probability for two groups of scientists. The first group includes the scientists who started their careers between 1950 and 1960, while the second group contains the scientists who started their careers between 1970 and 1980. One can see in Fig. 4c that the distributions of number of communities for these two groups of scientists largely overlap. The distributions of the switching probability for these two groups of scientists in Fig. 4d, however, exhibit significant difference.

We finally propose a model that could help to understand the main mechanisms leading to the observed patterns of scientists' research dynamics. The research activities of scientists can be modeled as discovery process in the knowledge space (i.e. a network characterizing the connections among different knowledges)~\cite{choosing2015rzhetsky,network2018iacopini}. When a scientist publishes a paper, she activates a node (i.e. a new knowledge) in the knowledge space. The sub-network activated by this scientist during her career forms a personal network recording all her papers as well as the links, i.e., relations between them. The simplest model for the node activation process is the standard random walk, assuming that a scientist randomly activates a neighboring node of the former activated node. Here, we propose an Exploitation-Exploration model (EEM) by introducing an exploitation process (controlled by a probability $p$) and an exploration process (controlled by a probability $q$) to the random walk model. Both processes have been pointed out to be fundamental for innovation in various adaptive systems~\cite{exploration1991march}. In our model, these two processes are performed sequentially. Instead of always starting from the last activated node in each step, the scientist has probability $p$ to randomly restart from (re-exploit) one of the previously activated nodes. Once the re-exploited node is determined, the scientist has probability $q$ to explore nodes beyond the nearest neighbors. For simplicity, we assume that the scientist randomly activates in the exploration step a next-nearest neighbor. Note that the EEM reduces to the standard random walk model when $p=0$ and $q=0$. For an illustrative demonstration of the random walk model and the EEM, see Fig. 5a. In our simulation, the knowledge space is represented as a network consisting of all the APS papers, with any two nodes (papers) linked if they share at least one reference. The first activated node for each scientist is set to be her first paper. The rest of the papers of each scientist are generated by following the EEM on the APS network until the number of activated nodes equals to the real number of papers of each scientist.

We first test the EEM by simulating the research dynamics of the representative highly-cited scientist presented in Fig. 1. Specifically, we compare in Fig. 5b the co-citing network (CCN) as well as the time series of published papers generated by both, the standard random walk model and the EEM. The the initial paper and the number of papers in each year of this simulated scientist are set the same as in the real data. One can immediately see that the network generated applying the standard random walk model is very different from the typical real one in Fig. 1b as it contains many long chains and it lacks distinct communities. Moreover, the time series obtained from the random walk model is also very different from that of typical real researcher shown in Fig. 1d in the sense that no switching between communities can be observed in each year. In contrast, both the network and the time series generated by the EEM qualitatively reproduce similar properties as those exhibited in Fig. 1. We further support quantitatively the EEM by examining some statistical quantities generated by this model. The first relates to the number of yearly involved communities under different $p$, as presented in Fig. 5c. When $p=0$, each scientist roughly works in only one community each year. As $p$ increases, the number of yearly involved communities becomes larger, with $p=0.6$ peaking around 1.8 which is the value observed in real data. We have tested and found that $q$ has little effect on the yearly involved communities, thus it is set to be $0$ in Fig. 5c. Another statistical quantity is the number of communities that each scientist is involved during her research career. When $q=0$, the generated sub-network does not have distinct communities and thus the number of communities is very narrowly distributed (even for $size>0$ case where all detected clusters are regarded as communities), as shown in Fig. 5d. As $q$ increases, small communities start to emerge, resulting in the separation of the distributions of the $size>0$, $size>2$ and $size>5$ cases. When $q=0.2$, the distributions of $size>0$, $size>2$ and $size>5$ cases respectively peak around 11, 8 and 5, similar to that in real data, see Fig. 3a. We have also found that $p$ has a little effect on the distribution of the number of communities, thus we set $p=0$ for Fig. 5d.

We finally estimate the probability $p$ and $q$ for each scientist based on real data. We denote the number of papers published by a scientist $i$ as $n_i$. When each of these papers is published, if it shares no reference with any of $i$'s papers published before, we keep a record of this paper and finally denote $u_i$ as the total number of such papers. $q_i$ then can be easily estimated as $q_i=u_i/n_i$. In the sequence of the $i$'s papers, if a paper shares at least one reference with the former paper published by $i$, we keep a record of this paper and finally denote $v_i$ as the total number of such papers. In this way, we can estimate $p_i$ as $p_i=v_i/(n_i-u_i)$. The distributions of the estimated $p$ and $q$ from real data are shown in Figs. 5e and 5f. One can see that the distributions of $p$ and $q$ peaks around $0.6$ and $0.2$ respectively, which are the same as the values in Figs. 5c and 5d that generate consistent statistical properties with real data.

\section*{Discussion}
To summarize, we study the research dynamics of scientists by constructing a network of each individual scientist's publications characterizing their co-citing relations. We find that typically each network appears to have a clear community structure. The papers in a community tend to share the same PACS code, indicating that each community indeed represents a research area. By filtering out the small communities of less than 3 nodes, we obtain the major communities of scientists. We find that the numbers of major communities of scientists during their career are narrowly distributed. In addition, the largest three communities already comprise over 70\% of most scientist's papers. We compare the statistical properties of the co-citing networks of scientists who started their career in different years. We find that though the total number of communities stays almost unchanged, yet the switching between communities tends to increase and becomes more frequent during the years. In addition, we find that high average citation per paper in all career stages correlates with low switching probability. In marked contrast, high switching probability in early career correlates with low overall productivity, while high switching probability in latter career is associated with high overall productivity. Finally, we propose a model capturing the main features of the research dynamics of individual scientists. The research activity is modeled as a node activation process in a knowledge network where nodes represent all the papers in APS and links represent co-reference relations between nodes. The model reproduces the main structural and dynamical patterns of individual scientist's publishing behavior by assuming the scientist activates nodes in the network based on a random walk process which includes the exploitation and exploration mechanisms.

Our work provides a general framework for incorporating network tools into the temporal analysis of publication records of individuals. Several promising extensions can be built on this work. A straightforward one would be constructing papers' network for departments or institutions, which will help us to estimate the cooperativity behavior in the department. The higher-level research dynamics of these departments or institutions might be fundamentally different from the research dynamics at individual scientist level, the study of which will substantially deepen our understanding of how research activities are collectively organized. Similarly, one can investigate the networks characterizing relations among the papers published under the support of cooperative or individual research grants. The outcome of a research grant can thus be evaluated not only based on the number of papers but also be based on the actual research directions and the cooperation between scientists.

\section*{Materials and methods}
\textbf{Data.} In this paper, we analyze the publication data from all journals of American Physical Society (APS). The data contains 482,566 papers, ranging from year 1893 to year 2010. For the sake of author name disambiguation, we use the author name dataset provided by Sinatra et al. which is obtained with a comprehensive disambiguation process in the APS data~\cite{quantifying2016sinatra}. Eventually, a total number of 236,884 distinct authors are matched. We found and analyzed 3,420 authors with at least 50 papers, and 15,373 authors with at least 20 papers. Another set of data that we analyzed in the supplementary materials is the computer science data obtained by extracting scientists' profiles from online Web databases~\cite{extraction2008tang}. The data contains 1,712,433 authors and 2,092,356 paper, ranging from year 1948 to year 2014. The author names in this data are already disambiguated. We found and analyzed 9,818 authors in this data with at least 50 papers.\\
\textbf{Community detection.} The co-citing network of a scientist is constructed by linking two papers if they share at least one reference. For simplicity, we do not weight the links and only consider the topology of the network. The community structure of the network is detected with the fast unfolding algorithm~\cite{fast2008blondel} which is a heuristic method based on modularity optimization. The modularity function considered in this paper is defined as
\begin{equation}
Q=\frac{1}{2m}\sum_{i,j}[A_{ij}-\gamma\frac{k_ik_j}{2m}]\delta(c_i,c_j),
\end{equation}
where $A$ is the adjacency matrix of the network, $k_i$ is the degree of node $i$, $m$ is the total number of links in the network, $c_i$ is the community to which node $i$ is assigned, the $\delta$ function $\delta(c_i,c_j)$ is 1 if $c_i=c_j$, and 0 otherwise. The communities are obtained when the function $Q$ is maximized. Note that $\gamma$ is a resolution parameter in $Q$~\cite{statistical2006reichardt,limited2007kumpula}, with $\gamma=1$ in the standard modularity function~\cite{fast2004newman}. A larger $\gamma$ results in detecting small but more communities, while a smaller $\gamma$ yields larger but fewer communities. Results with $\gamma\neq1$ are presented in the supplementary materials. Although the distribution of the number of communities is influenced by the parameter $\gamma$ (see Fig. S12), the dynamics properties are shown to be almost independent of the resolution of communities (see Fig. S13). For this reason, we consider the standard modularity function, i.e. $\gamma=1$, in this paper.

\bibliography{scibib}

\bibliographystyle{Science}

\section*{Acknowledgments}
We thank Junming Huang and and Louis Shekhtman for useful discussions. This work is supported by the National Natural Science Foundation of China (Grant Nos. 61603046, 61773069, 71731002 and 61573065) and the Natural Science Foundation of Beijing (Grant No. L160008). ZS is supported by China Postdoctoral Science Foundation under Grant 2017 M620944. HES acknowledges the support from NSF Grants PHY-1505000, CMMI-1125290, and CHE-1213217, and DTRA Grant HDTRA1-14-1-0017. SH acknowledges the Israel-Italian collaborative project NECST, the Israel Science Foundation, U.S. Army Research Office contract number W911NF1810396, ONR, the Israeli Most and Japan Science Foundation, BSF-NSF, and DTRA (Grant No. HDTRA-1-10-1-0014) for financial support.

\section*{Author contributions}
AZ, YW, HES, SH designed the research, AZ, ZS and JZ performed the experiments, AZ, YW and SH analyzed the data, all authors wrote the manuscript.\\

\section*{Competing financial interests} The authors declare no competing financial interests.\\

\section*{Data and materials availability} The data used in this paper are all publicly accessible. The APS data can be downloaded via \url{https://journals.aps.org/datasets}, and the computer science data can be downloaded via \url{https://www.aminer.cn/aminernetwork}.

\section*{Supplementary materials}
Figs. S1 to S16\\

\clearpage
\section*{Figures}
\begin{figure}[h!]
  \centering
  \includegraphics[width=16cm]{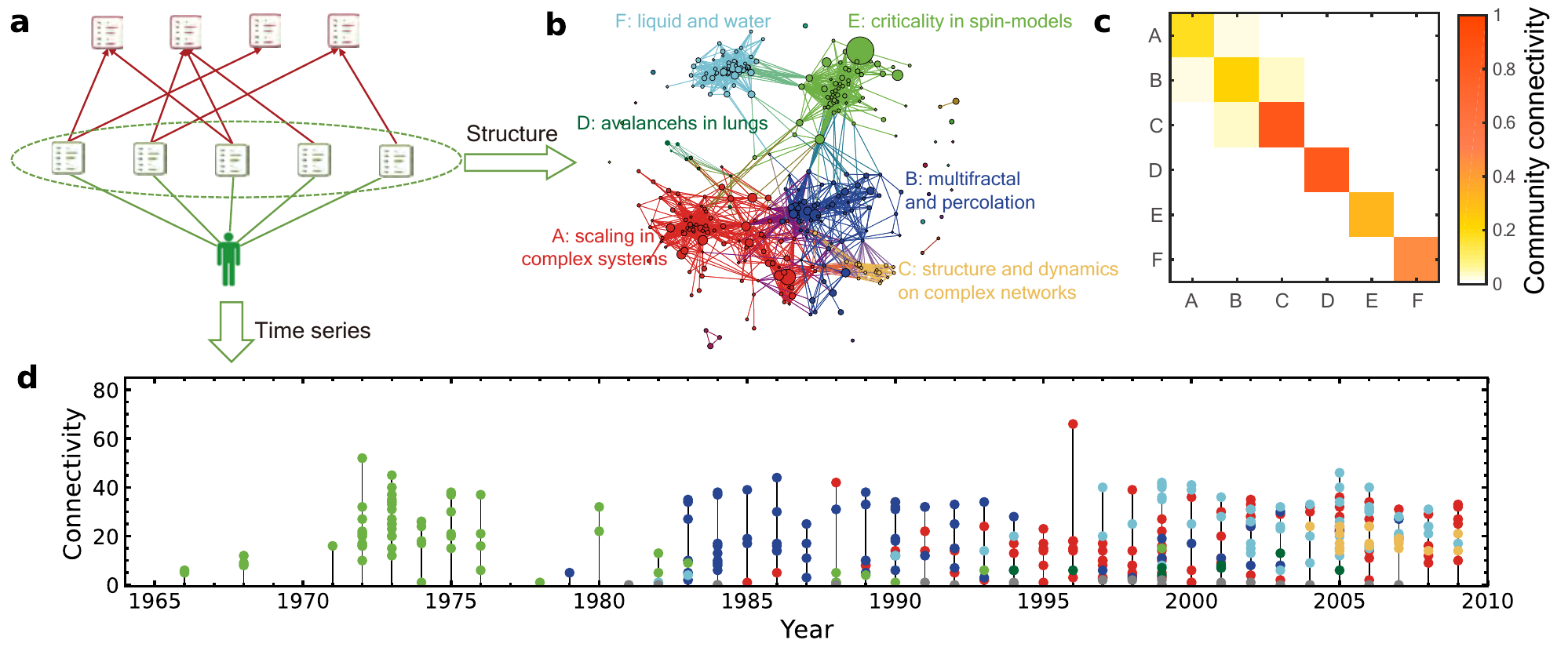}\\
  \caption{Illustration of the co-citing network (CCN) of a typical highly-cited scientist and its growth history. (a) The data and method used to construct the co-citing network. The papers authored by the scientist are marked in green and the references of these papers are marked in red. (b) The co-citing network consists of all the papers published by this scientist. Each paper is represented by a node, and two papers are connected if they share at least one reference. The communities of this network are identified with the fast unfolding algorithm which detects communities by maximizing the modularity function. The network contains several large-size communities as well as some small clusters and isolated nodes. Each major community represents a main research topic of this scientist. (c) The community connectivity matrix shows that nodes within each community are well connected, yet nodes of different communities are much less connected. (d) The time series presented at the bottom describes the growth history of the network and meanwhile reveals how this scientist moves from one research topic to another during her career. In the sub-figure of time series, each point is a paper and the color corresponds to the community in the co-citing network. The height of the point is the number of links (i.e. connectivity) that the paper has in the network.}\label{fig1}
\end{figure}

\clearpage
\begin{figure}[h!]
\centering
	\includegraphics[width=15cm]{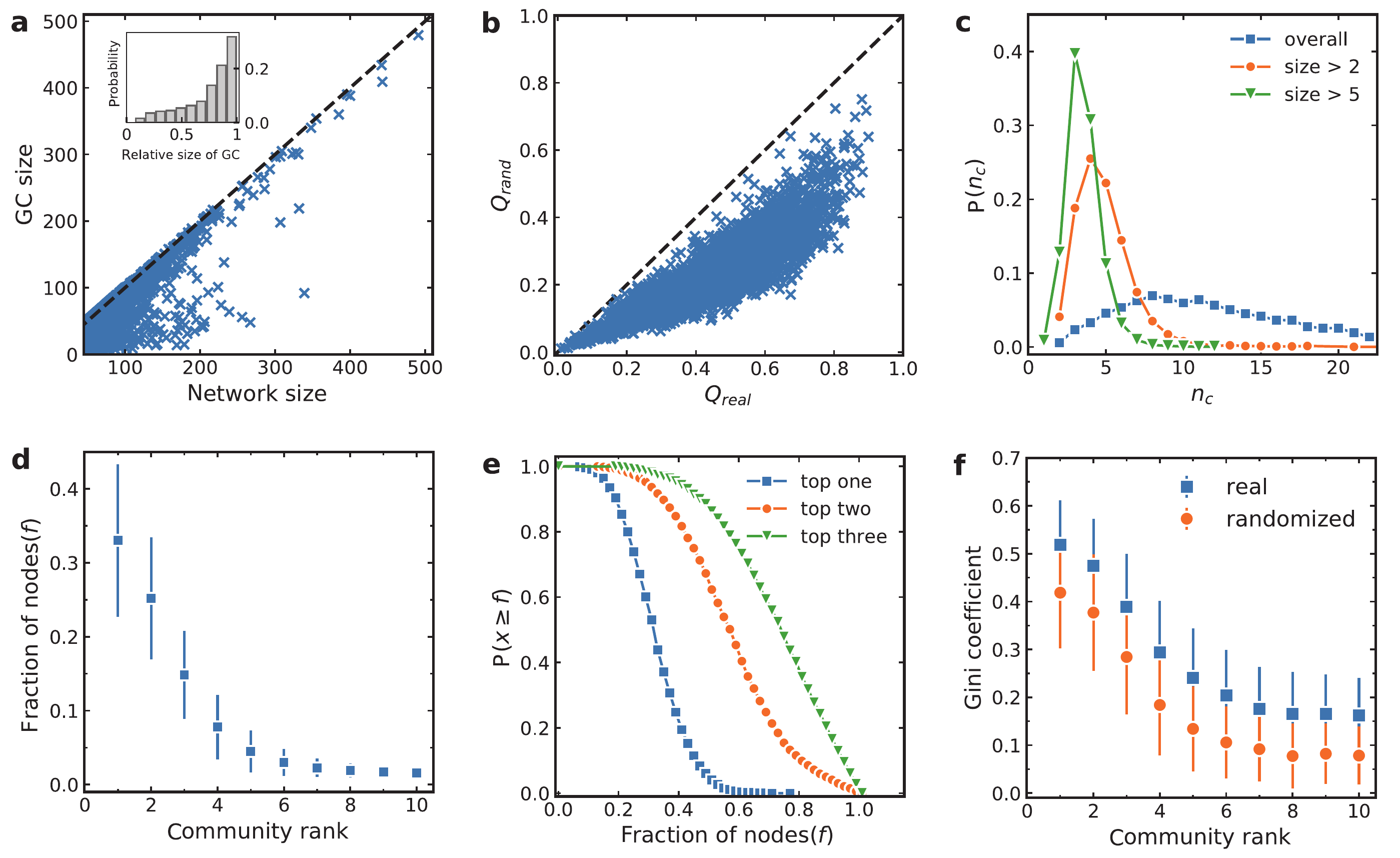}\\
	\caption{(a) The size of the co-citing network (CCN) versus the size of CCN's giant component (GC). Each point represents a scientist. Most of the points are located below but close to the diagonal line, indicating that CCNs are in general connected and have relatively large GCs. This is supported by the inset where the distribution of the relative size of GC is presented. (b) The maximized modularity in real CCNs and the maximized modularity in their degree-preserved reshuffled counterparts. All the points are located under the diagonal line, indicating that the community structure in real networks is truly significant. (c) The distribution of the number of communities for all scientists. Three curves are presented where all communities are taken into account (legend as ``all communities"), small communities with less than 3 nodes are eliminated (legend as ``size$>$2"), and small communities with less than 6 nodes are eliminated (legend as ``size$>$5"). (d) Fraction of papers in different communities. (e) Inverse cumulative probability of fraction of nodes in the biggest community (legend as ``top one"), the two largest communities (legend as ``top two"), and the three largest communities (legend as ``top three"), respectively. (f) The Gini coefficient of the distribution of PACS codes in different communities. A larger Gini coefficient corresponds to a more heterogeneous distribution, suggesting that higher fraction of papers in a community share the same PACS codes. The real data is compared with a random counterpart where the PACS codes are reshuffled among each individual scientist's papers while the community structure is preserved.}\label{fig2}
\end{figure}

\clearpage
\begin{figure}[h!]
  \centering
  \includegraphics[width=11cm]{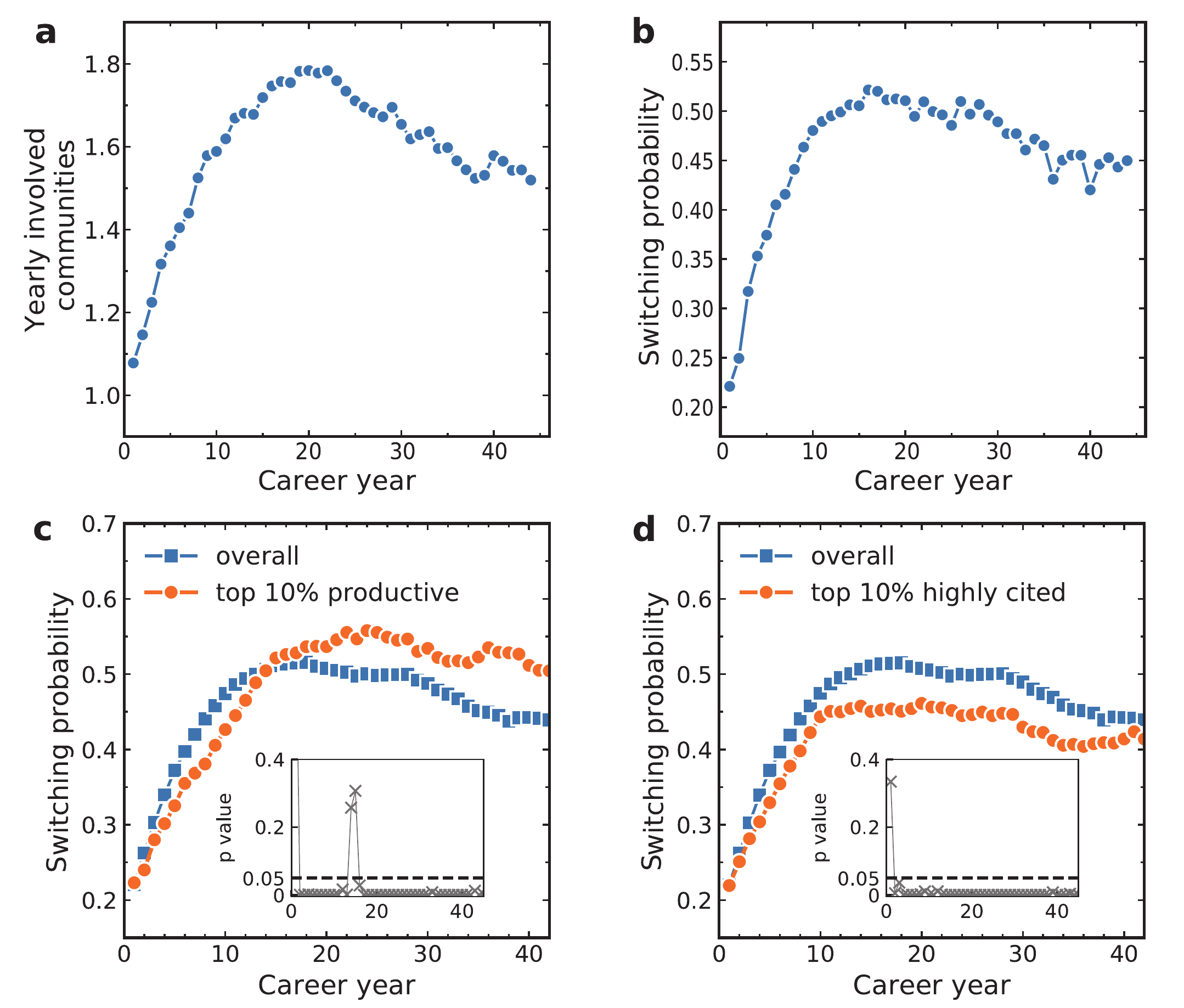}\\
  \caption{(a) The mean number of yearly involved major communities for individual scientists in different career years. (b) The switching probability between two adjacent publications from one major community to another major community of scientists in different career years. (c) Comparison of the overall switching probability (all scientists) with the switching probability of the 10\% most productive scientists in different career years. The results suggest that high productivity is associated with low switching probability in the early career but with high switching probability in the later career. (d) Comparison of the overall switching probability (all scientists) with the switching probability of the 10\% scientists who has the highest mean citation per paper. For each paper, we only consider the number of citations 10 years after its publication ($c_{10}$)~\cite{quantifying2016sinatra}. The results suggest that high average citation per paper in all career periods correlates with low switching probability. In the insets of (c)(d), we present the p-value of the Kolmogorov-Smirnov test distinguishing between the two switching probability distributions in each career year.}\label{fig3}
\end{figure}

\clearpage
\begin{figure}[h!]
  \centering
  \includegraphics[width=11cm]{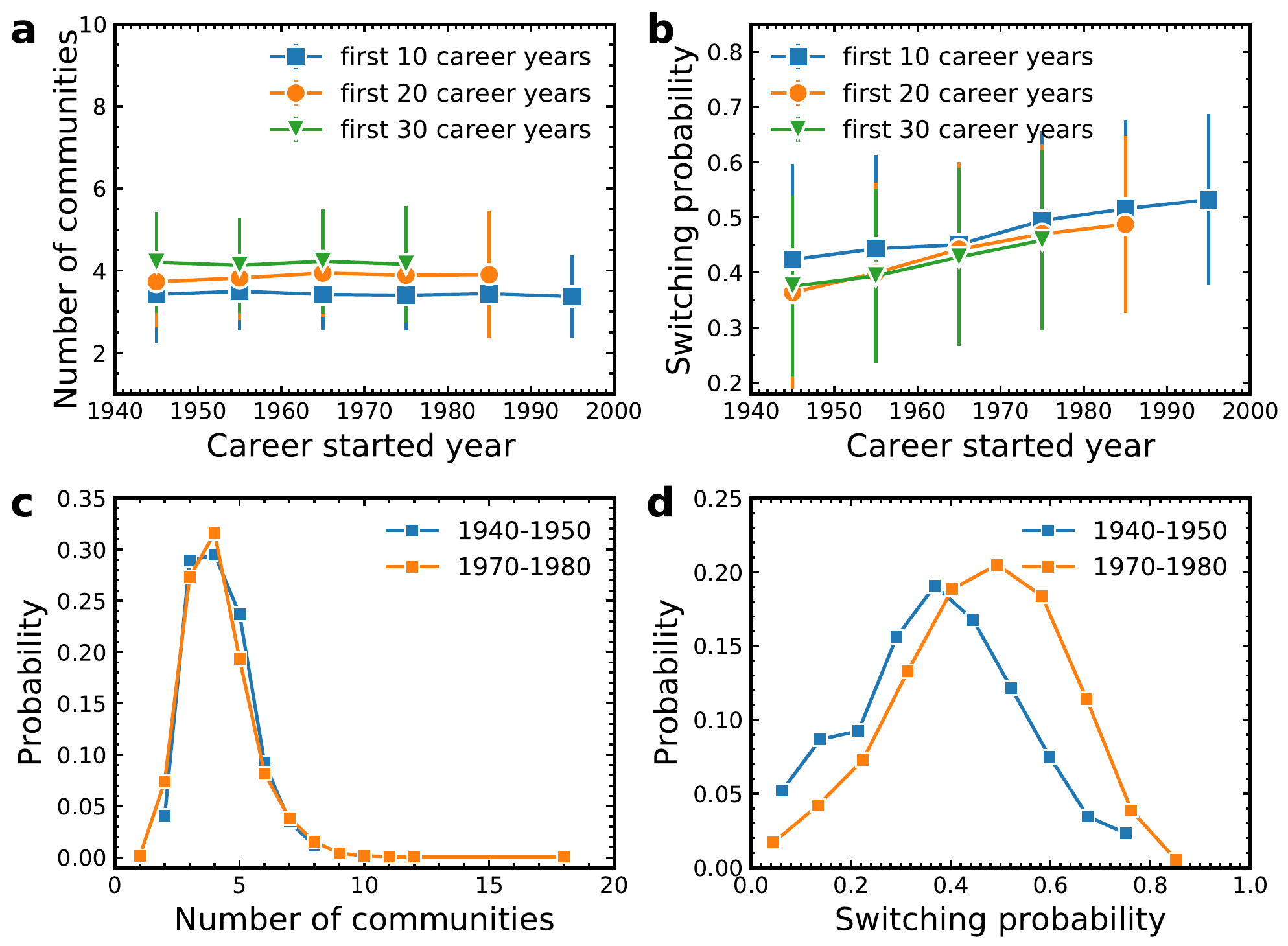}\\
  \caption{(a) The mean number of communities of scientists who started their career in different years. (b) The average switching probability of scientists who started their career in different years. As our data ends in 2010, it cannot capture the full career of scientists who started their careers in recent years. We thus filter out some scientists when we study the evolution of science here. We only consider scientists' first $y$ career years and remove (i) all the scientists that did not reach yet $y$ years of career (for a fair temporal comparison), and (ii) those who published less than 30 papers in their first $y$ career years (for a meaningful community detection). The results of $y=10, 20, 30$ are presented in this figure. As science evolves (during the years), the major communities that each scientist has stays almost unchanged, while the frequency that scientists switch between communities increases during the years. (c) Distributions of the number of communities (for $y=30$) for scientists who started their career between 1940 and 1950, and for those who started their career between 1970 and 1980. The p-value of the Kolmogorov-Smirnov test is $0.961$, suggesting significant similarity between these two distributions. (d) Distributions of the switching probability (for $y=30$) of scientists who started their career between 1940 and 1950, and of those who started their career between 1970 and 1980. The p-value of the Kolmogorov-Smirnov test is $ 2.34\times10^{-8}$, suggesting a significant difference between these two distributions (i.e., increase of switching probability).}\label{fig4}
\end{figure}

\clearpage
\begin{center}
  \centering
  \includegraphics[width=16cm]{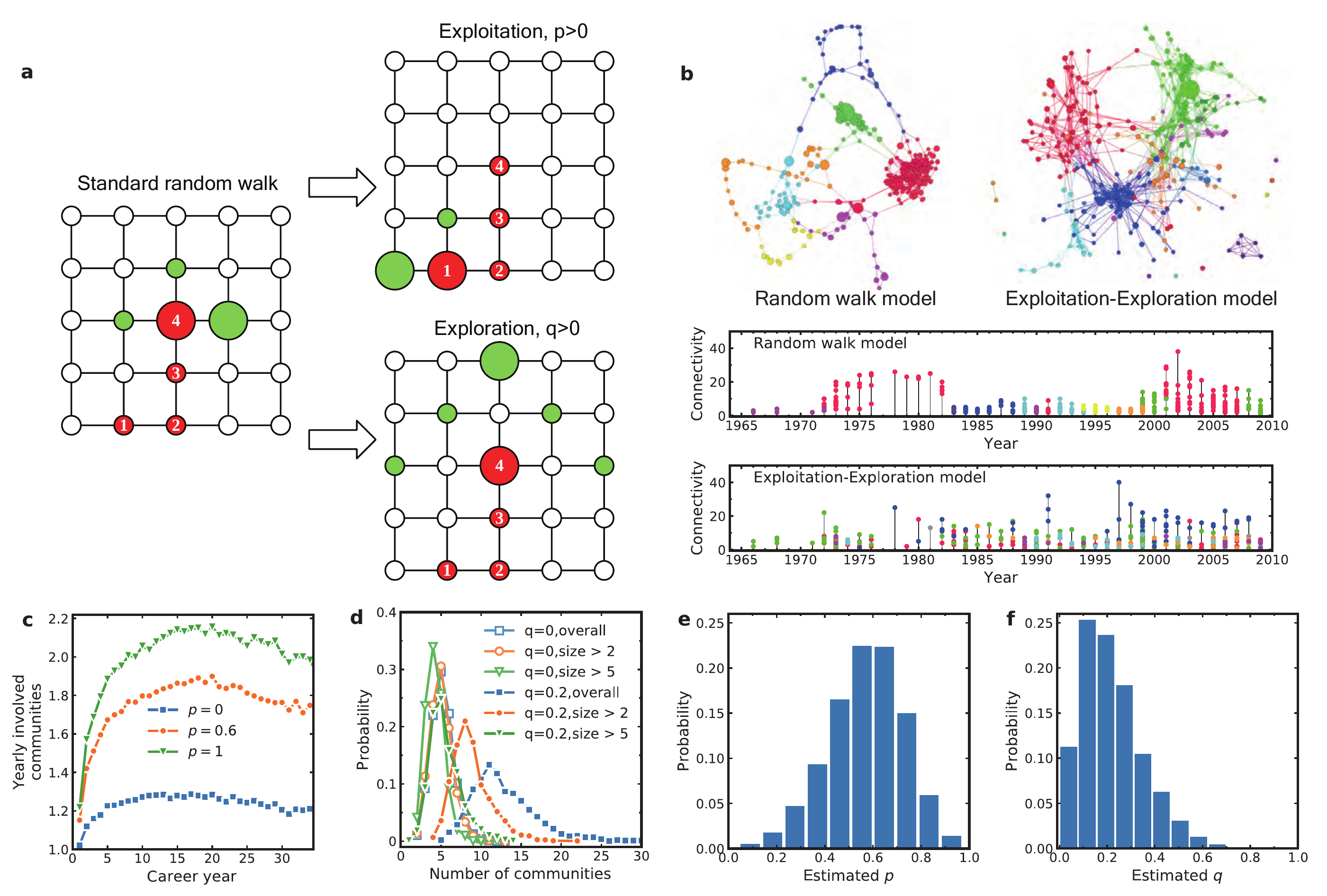}\\
  	\bigskip
	\setbox0\vbox{\makeatletter
		\let\caption@rule\relax
\captionof{figure}[short caption]{(a) Illustration of the Exploitation-Exploration model (EEM). The research activity is modeled as a node activation process in the knowledge space. When a scientist publishes a paper, she activates a node (i.e. a new knowledge) in the knowledge space. The network activated by this scientist at the end forms her personal network recording all her papers and the relations between them. The underlying lattice is a representation of the knowledge space for demonstration and the red nodes are the nodes already activated by a scientist, with a number recording the step in which the node is activated. The node activated in a later step is marked with a bigger size. The simplest model for the node activation process is the standard random walk, assuming that a scientist randomly activates a neighboring node of the previous activated node. Therefore, one of the neighboring nodes (marked in green) of the red node $4$ will be randomly picked and activated. The picked green node is marked with a bigger size. In the EEM, we introduce both, an exploitation process and an exploration process. With probability $p$, the scientist randomly re-exploits the neighborhood of one of the previously activated nodes. In the figure, the scientist makes exploitation by jumping back to the red node $1$ and randomly activating one of its neighbors. With probability $q$, the scientist explores nodes beyond the closest neighbors. For simplicity, we assume that a scientist randomly activates a next-nearest neighbor from the last activated node if $p=0$, or a next-nearest neighbor of the re-exploited node if $p>0$. In the figure, we assume $p=0$ and the scientist performs exploration by randomly activating a node among the next nearest neighbors of red node $4$. (b) Comparison of the co-citing networks (CCN) as well as the paper publishing time series generated by the random walk model and by the EEM. Note that the lattice is used here only for demonstration of the model. In our simulation, the knowledge space is characterized as the network consisting of all the APS papers, with any two nodes (papers) linked if they share at least one reference. The parameters including the initial paper and the number of papers in each year are set the same as in Fig. 1. In (c) and (d) these parameters are of all analyzed authors. (c) The number of yearly involved communities for different $p$, while $q=0$. (d) The distribution of the number of communities that each scientist is involved during her career for different $q$. (e)(f) Estimation of the probability $p$ and $q$ of each scientist based on the real data, plotted are their probability density functions.}
\global\skip1\lastskip\unskip
		\global\setbox1\lastbox
	}
	\unvbox0
	\setbox0\hbox{\unhbox1\unskip\unskip\unpenalty
		\global\setbox1\lastbox}
	\unvbox1
	\vskip\skip1
\end{center}\label{fig5}

\clearpage

\begin{center}
{\large\bfseries Supplementary Information}\\[8pt]
{\large Increasing trend of scientists to switch between topics}\\[8pt]
\small An Zeng, Zhesi Shen, Jianlin Zhou, Ying Fan, Zengru Di,\\
\small Yougui Wang, H. Eugene Stanley, and Shlomo Havlin\\
\end{center}


\begin{figure}[h!]
  \centering
  \includegraphics[width=11cm]{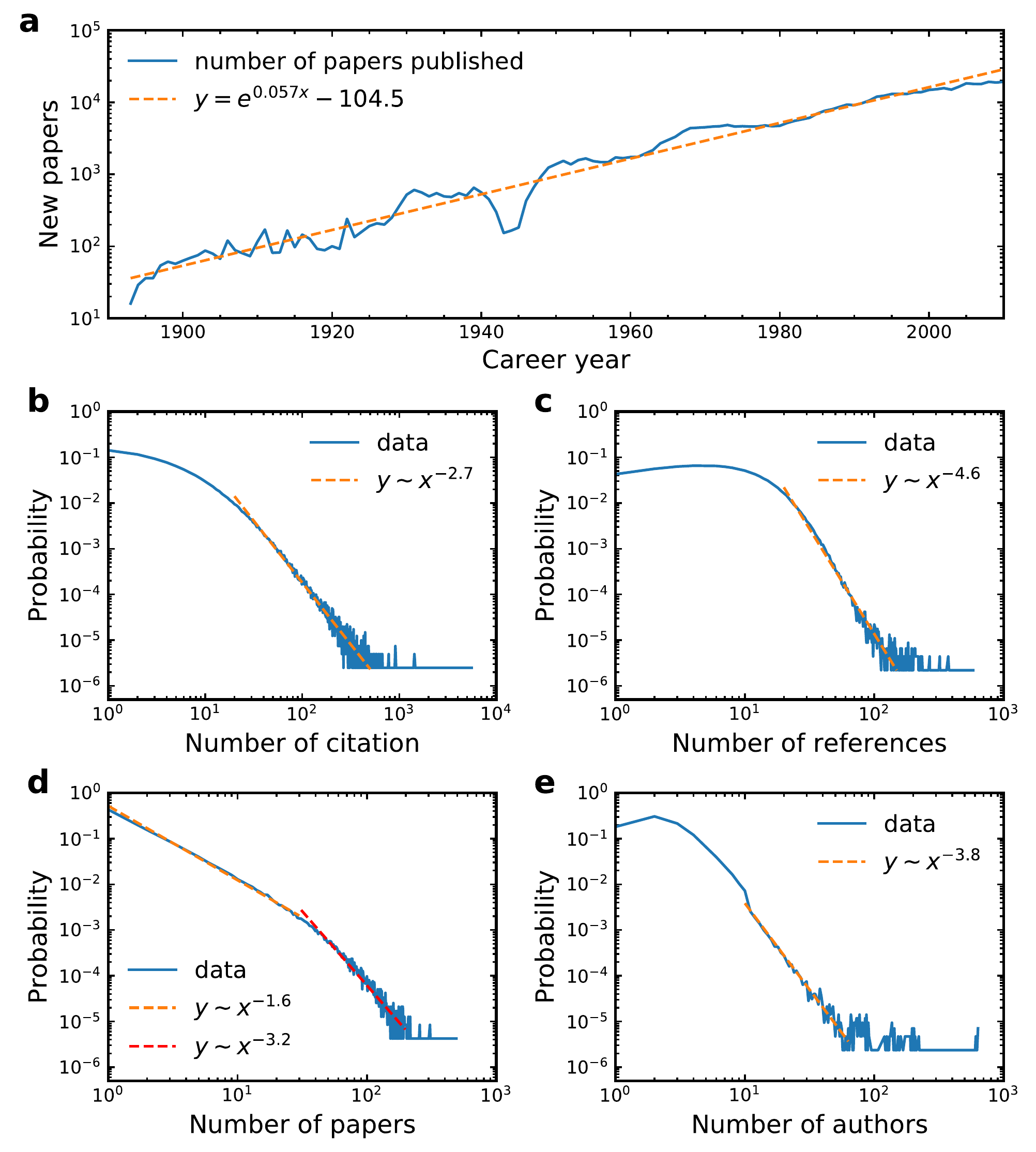}\\
  \textbf{Figure S1.} (a) Exponential fit to the yearly growth of new papers in the APS data set, with the trend consistent with that in the literature~\cite{centuary2015sinatra}. (b) Power-law fit to papers' citation distribution. (c) Power-law fit to papers' number of reference distribution. (d) Two power-law fits to authors' productivity (i.e. number of published papers of an author) distribution, in different regimes. (e) Power-law fit to papers' team size (i.e. number of authors) distribution. The fat long tail in (e) is due to papers from experimental nuclear and particle physics that have a huge number of authors.\label{FigS1}
\end{figure}

\begin{figure}[h!]
  \centering
  \includegraphics[width=15cm]{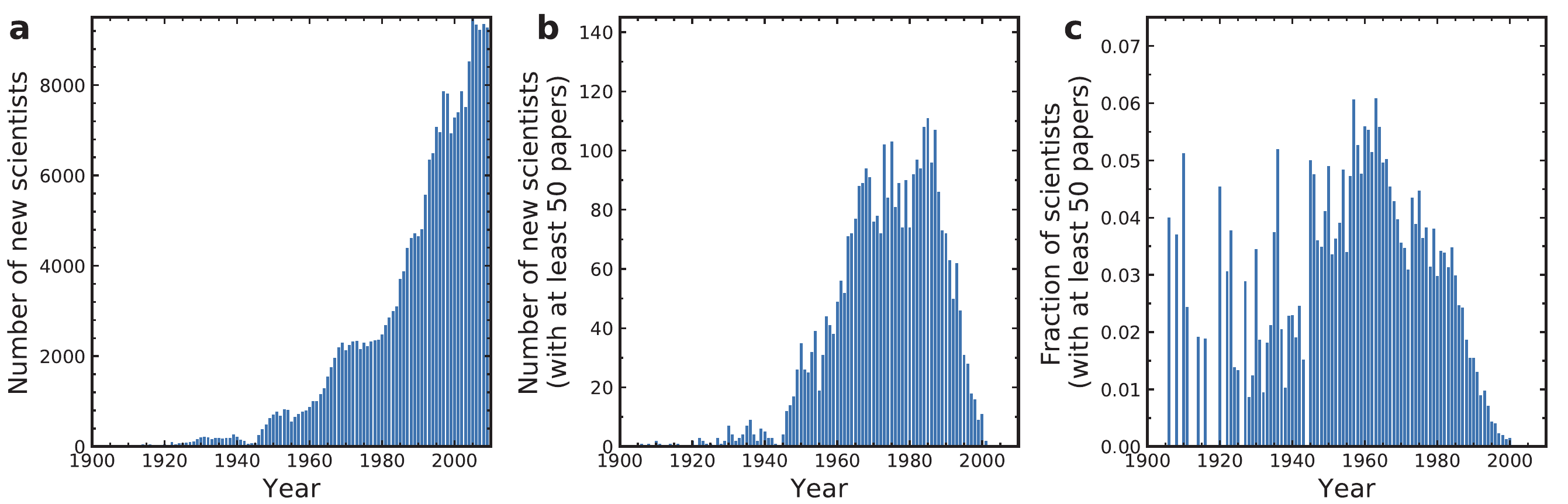}\\
  \textbf{Figure S2.} (a) The yearly number of new authors in APS data set. (b) The yearly number of new authors who will have at least 50 papers in APS data set. (c) The yearly fraction of new authors who will have at least 50 papers in APS data set. This fraction is generally stable over time, with a decreasing trend after 1990. This is because our data set ends in 2010 and the scientists who started their career after 1990 do not have enough time in the data set to cumulate at least 50 papers.\label{FigS2}
\end{figure}

\begin{figure}[h!]
  \centering
  \includegraphics[width=11cm]{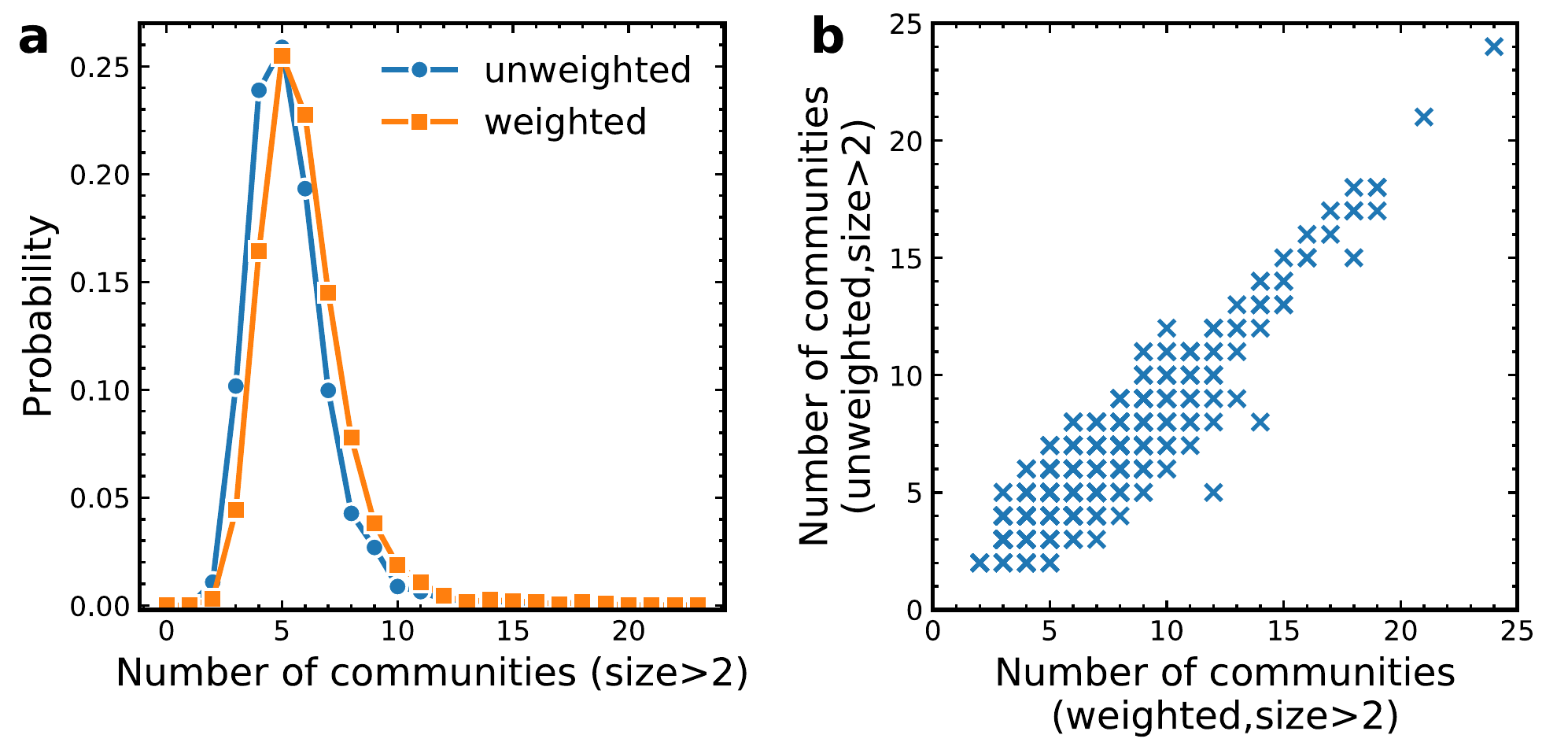}\\
   \textbf{Figure S3.} (a) The distributions of the number of communities in scientists' weighted and unweighted co-citing networks. Small clusters with no more than 2 nodes are filtered. In the weighted co-citing networks, the weight of a link is the number of common references between two connected papers. The communities in the weighted co-citing networks are obtained by maximizing the weighted modularity function~\cite{fast2008blondel}. (b) The scatter plot of the number of communities in scientists' weighted co-citing networks and the number of communities in their unweighted co-citing networks. The community structure is not significantly altered by considering weights, as links within communities tend to have large weights.\label{FigS3}
\end{figure}

\begin{figure}[h!]
  \centering
  \includegraphics[width=11cm]{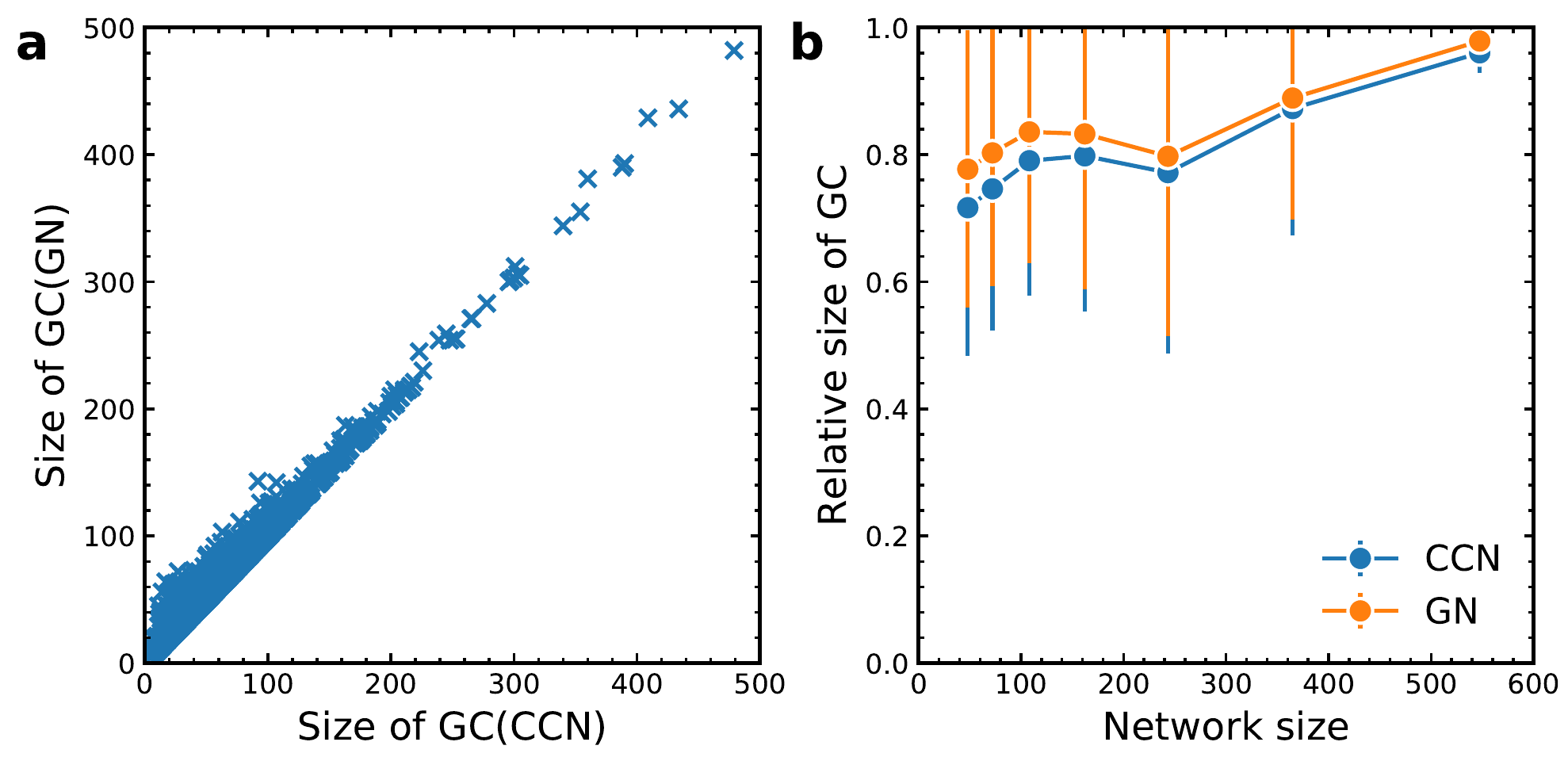}\\
  \textbf{Figure S4.} (a) For each scientist, we construct a co-citing network (CCN) based on the co-citing relations between her papers. We can also construct an aggregated network (GN) where any pair of her papers are linked if they either co-cite at least one reference or are co-cited by at least one paper. We compare the difference between CCN and GN by making a scatter plot of the size of the giant component in these two networks. Most of the points are located very near the diagonal, indicating that using co-citing relations can well capture the overall relations between papers. (b) The relative sizes of giant component (GC) of CCN and GN for scientists with different number of papers, showing that the GCs are generally a large fraction of the network for all scientists in both CCN and GN.\label{FigS4}
\end{figure}

\begin{figure}[h!]
  \centering
  \includegraphics[width=11cm]{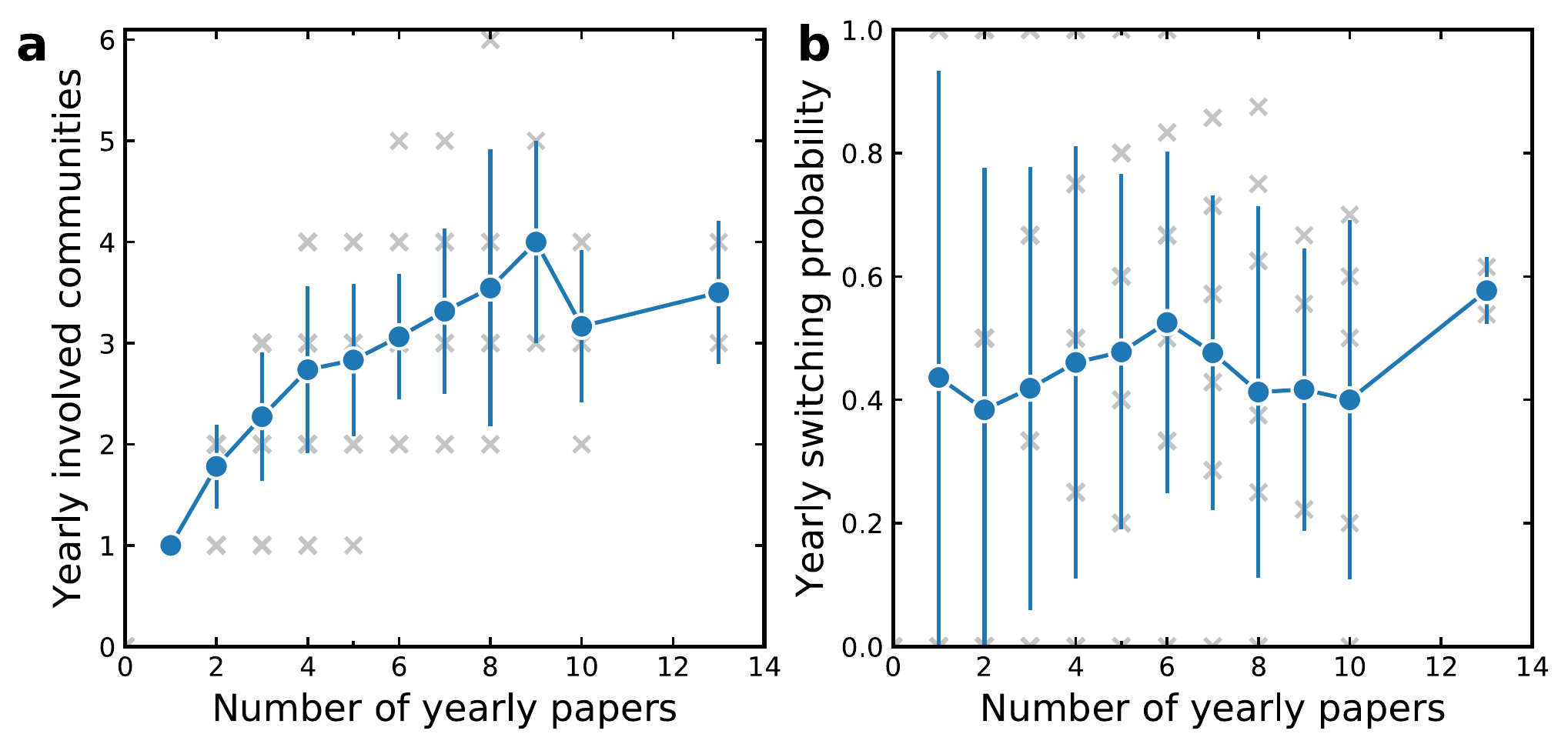}\\
  \textbf{Figure S5.} (a) The yearly involved communities versus the yearly published papers for all scientists in each of their first 40 career years. When a scientist publishes more papers in a year, he/she will have higher number of yearly involved communities purely by chance. (b) The yearly switching probability versus the yearly published papers for all scientists in each of their first 40 career years. The results demonstrate that the yearly switching probability is not correlated with the yearly published papers, and thus can be used as an unbiased metric for quantifying scientists' switching behavior between communities. \label{FigS5}
\end{figure}

\begin{figure}[h!]
  \centering
  \includegraphics[width=11cm]{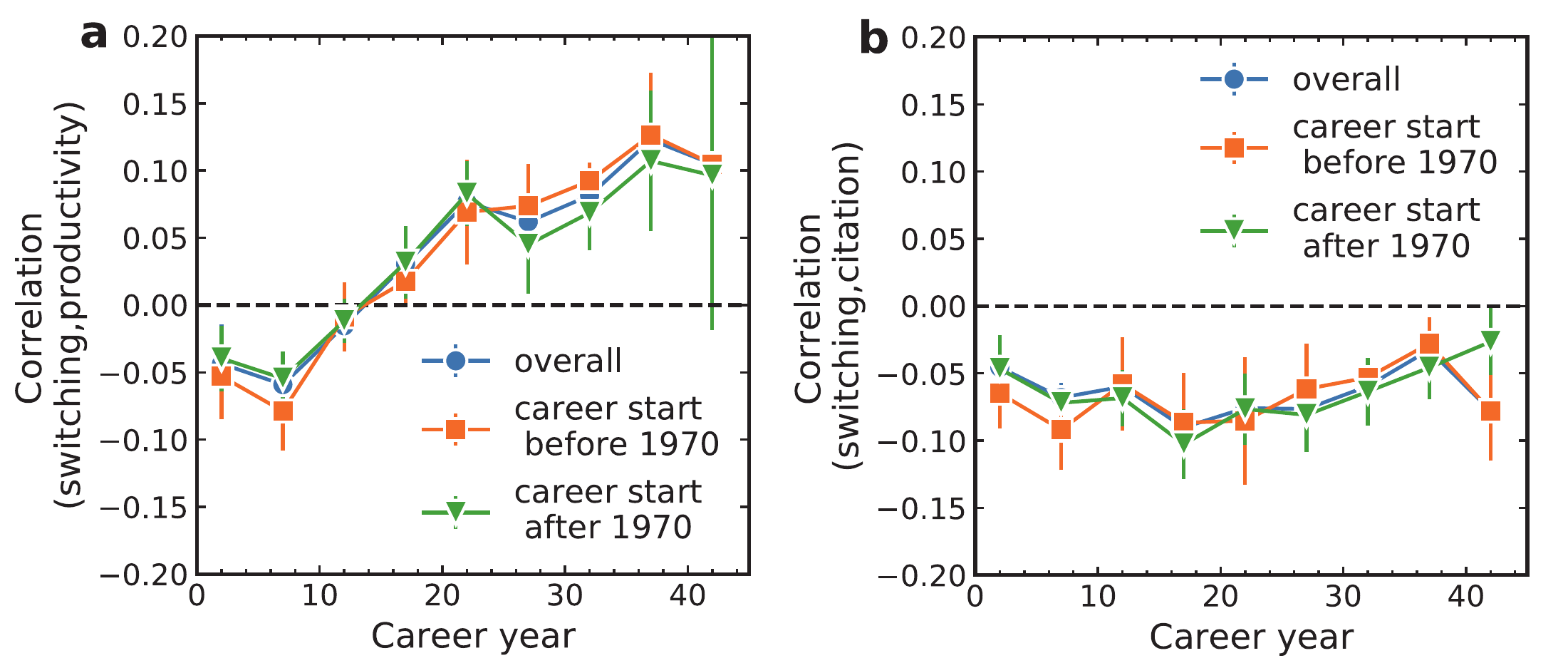}\\
   \textbf{Figure S6.} (a) The Pearson correlation between scientists' switching probability in different career years and their overall productivity. We find that high switching probability in early career is correlated to low overall productivity (negative correlations), while high switching probability in later career is associated with high overall productivity (positive correlations). (b) The Pearson correlation between scientists' switching probability in different career years and the mean citations per paper. The average citation per paper is negatively correlated with the switching probability in all career periods. The correlations are robust for scientists who start their careers in different years.
   \label{FigS6}
\end{figure}

\begin{figure}[h!]
  \centering
  \includegraphics[width=12cm]{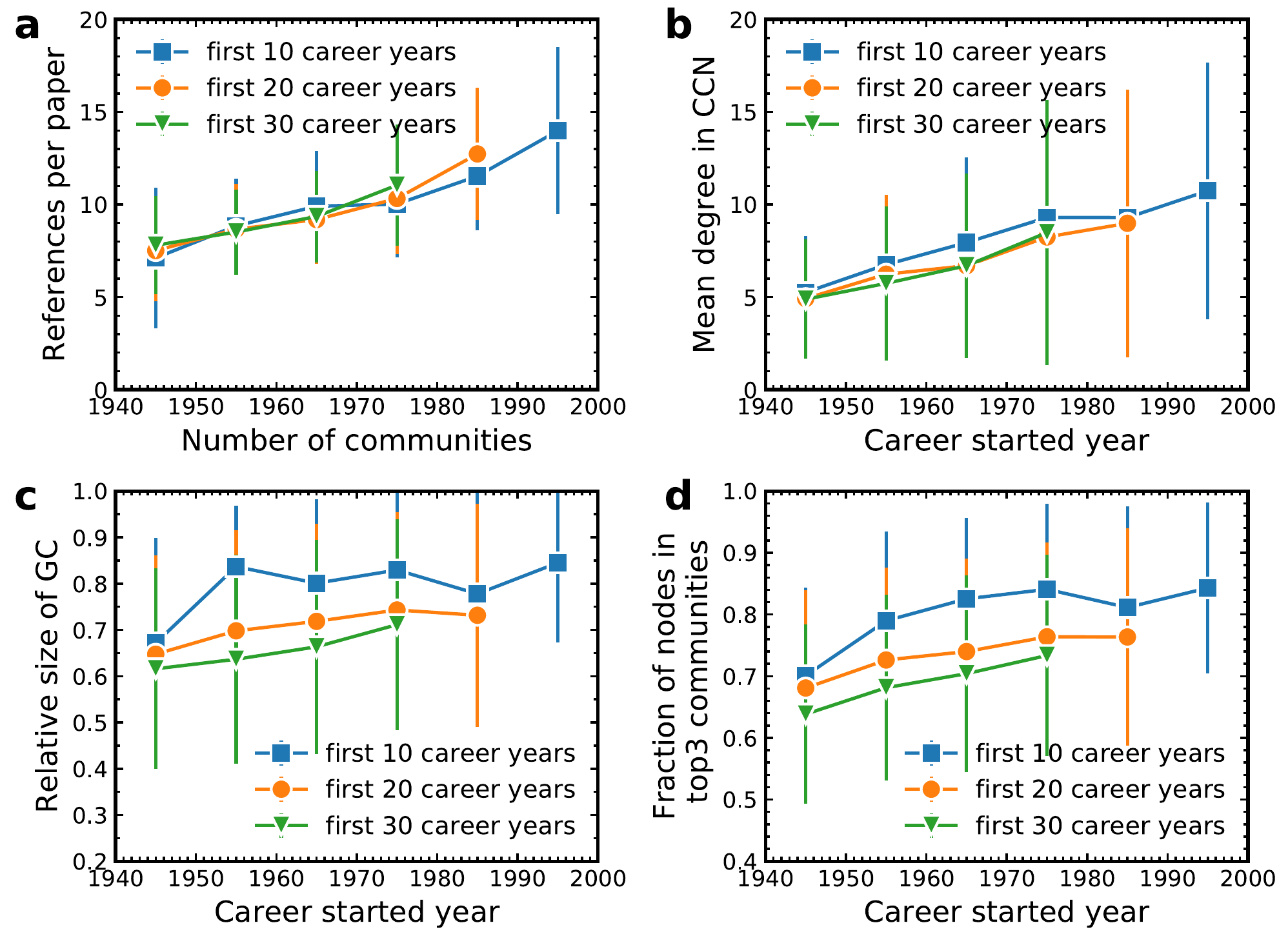}\\
  \textbf{Figure S7.} (a) The mean number of references per paper published by scientists who started their career in different years. (b) The mean degree of the co-citing networks (CCNs) of scientists who started their career in different years. (c) The mean relative sizes of GC of the CCNs of scientists who started their career in different years. (d) The mean fraction of nodes in the 3 largest communities in the CCNs of scientists who started their career in different years. \label{FigS7}
\end{figure}

\begin{figure}[h!]
  \centering
  \includegraphics[width=10cm]{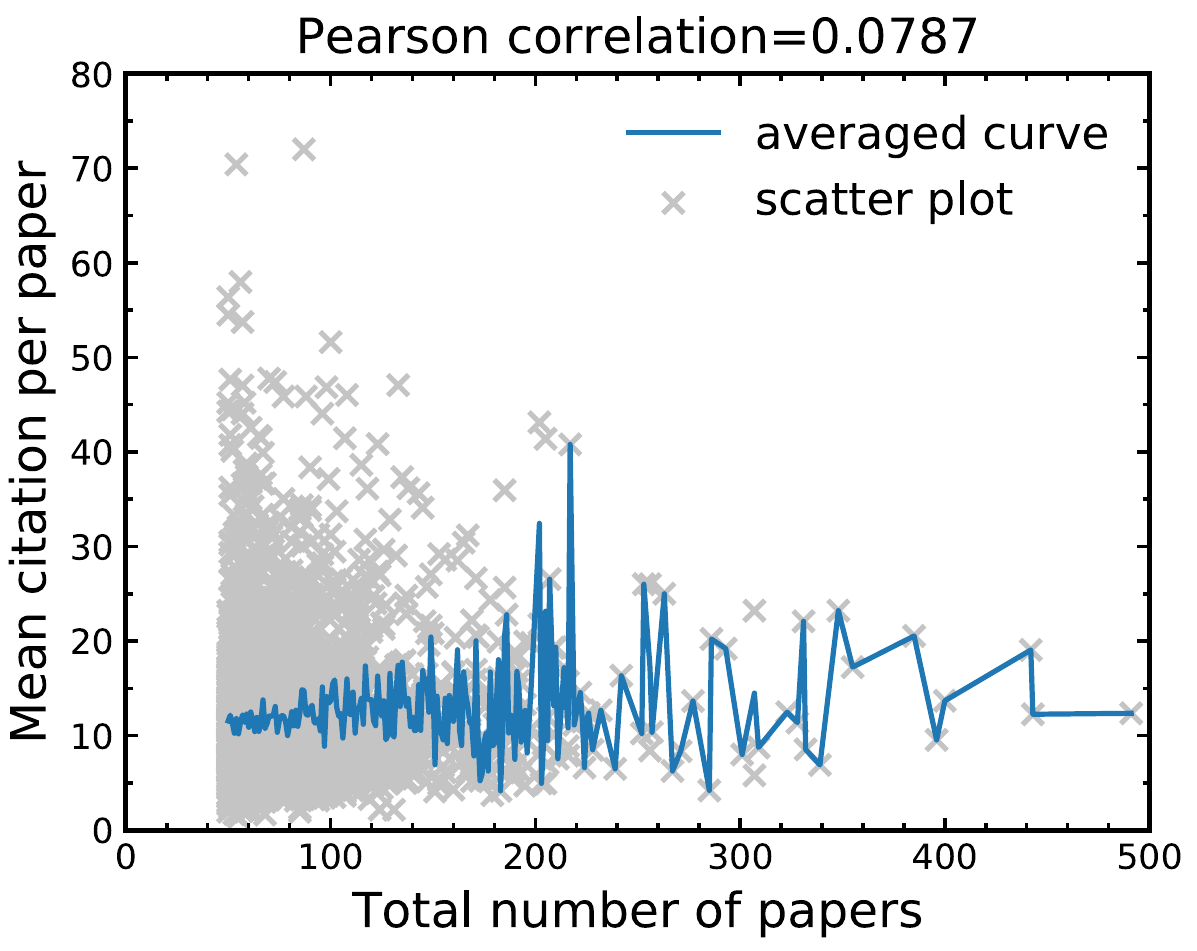}\\
  \textbf{Figure S8.} Scatter plot of the number of published papers and the mean citation per paper of each scientist. The blue curve shows the average trend, indicating that these two performance metrics are almost uncorrelated. \label{FigS8}
\end{figure}

\begin{figure}
  \centering
  \includegraphics[width=13cm]{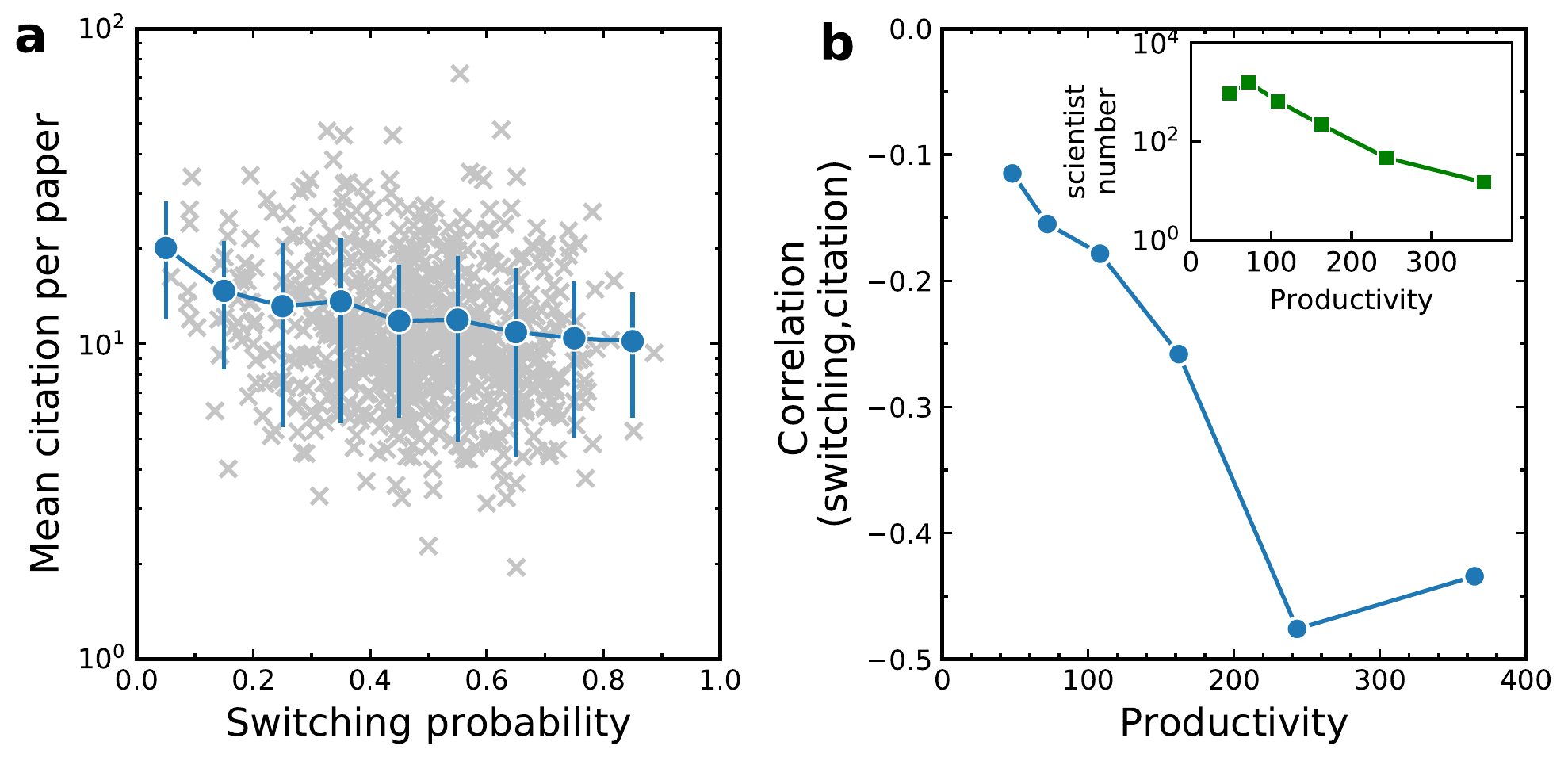}\\
  \textbf{Figure S9.} (a) The switching probability versus mean citation per paper for the scientists who published between 70 and 90 papers in their career. The downward trend shown by the averaged curve indicates that the switching probability and mean citation per paper is slightly negatively correlated. (b) The correlation between the switching probability and mean citations per paper for scientists of different productivity (number of papers in their career) range. The inset shows the number of scientists in each productivity range. One can see that the correlation between switching probability and mean citation per paper is negatively correlated for each group of scientists, and the negative correlation is more significant for more productive scientists. \label{FigS9}
\end{figure}

\begin{figure}[h!]
  \centering
  \includegraphics[width=16cm]{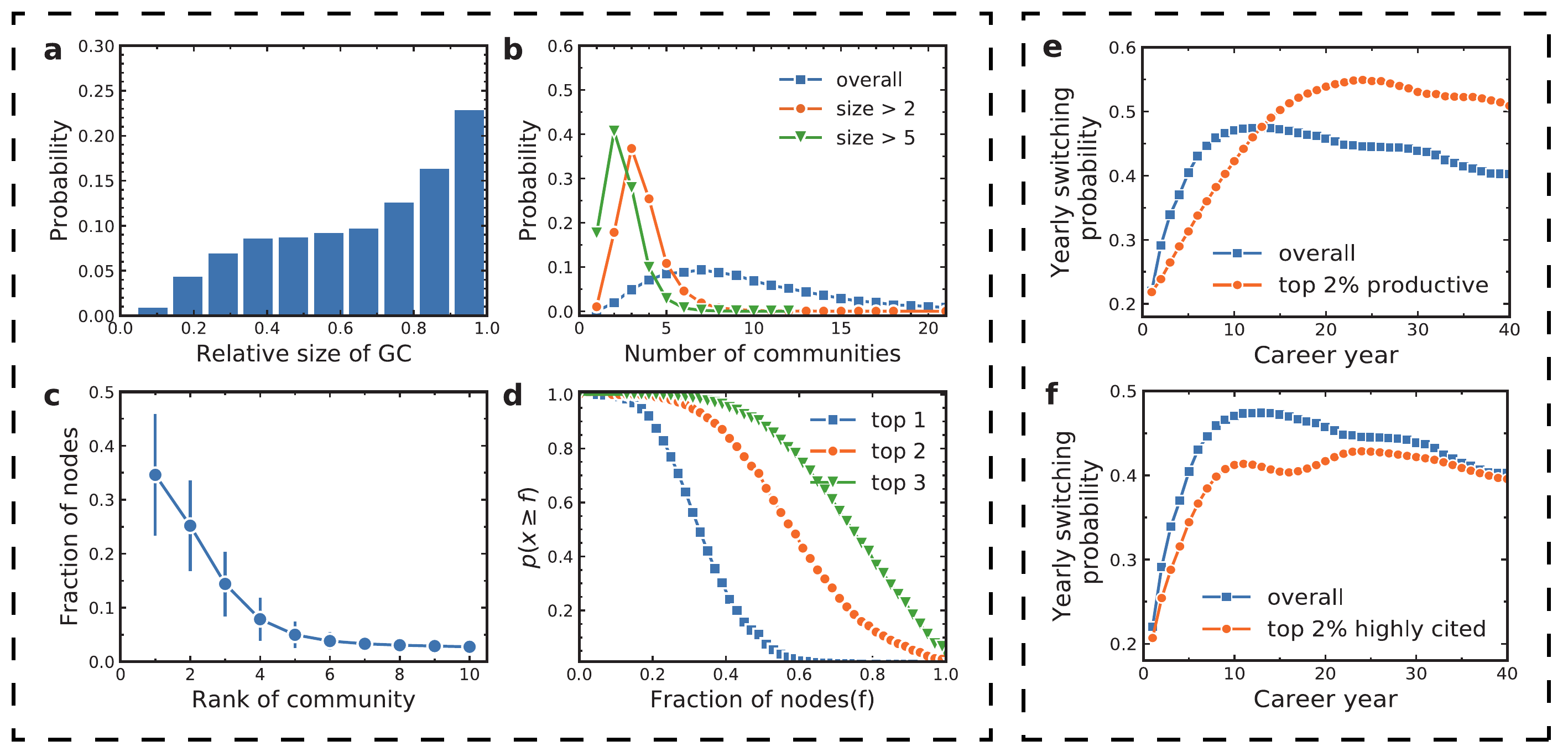}\\
  \textbf{Figure S10.} In the paper, we only consider scientists with at least 50 papers in order to ensure that the network is sufficiently large to obtain meaningful community detection results. Here, we test an alternative case where all the scientists with at least 20 papers are considered, which results in 15373 scientists in our data set. This figure shows the structural analysis results. (a) The distribution of the relative size of GC. (b) The distribution of the number of communities for all scientists. The three curves respectively represent the cases where all communities are preserved, small communities with fewer than 2 nodes are eliminated, and small communities with fewer than 5 nodes are eliminated. (c) Fraction of papers in different communities sorted by descending size. (d) Inverse cumulative probability of fraction of nodes in the biggest community (legend as ``top 1"), the two largest communities (legend as ``top 2"), and the three largest communities (legend as ``top 3"), respectively. (e) Comparison of the overall switching probability with the switching probability of the 2\% most productive scientists in different career years. (f) Comparison of the overall switching probability with the switching probability of the 2\% scientists who has the highest mean citation per paper. \label{FigS10}
\end{figure}

\begin{figure}[h!]
  \centering
  \includegraphics[width=13cm]{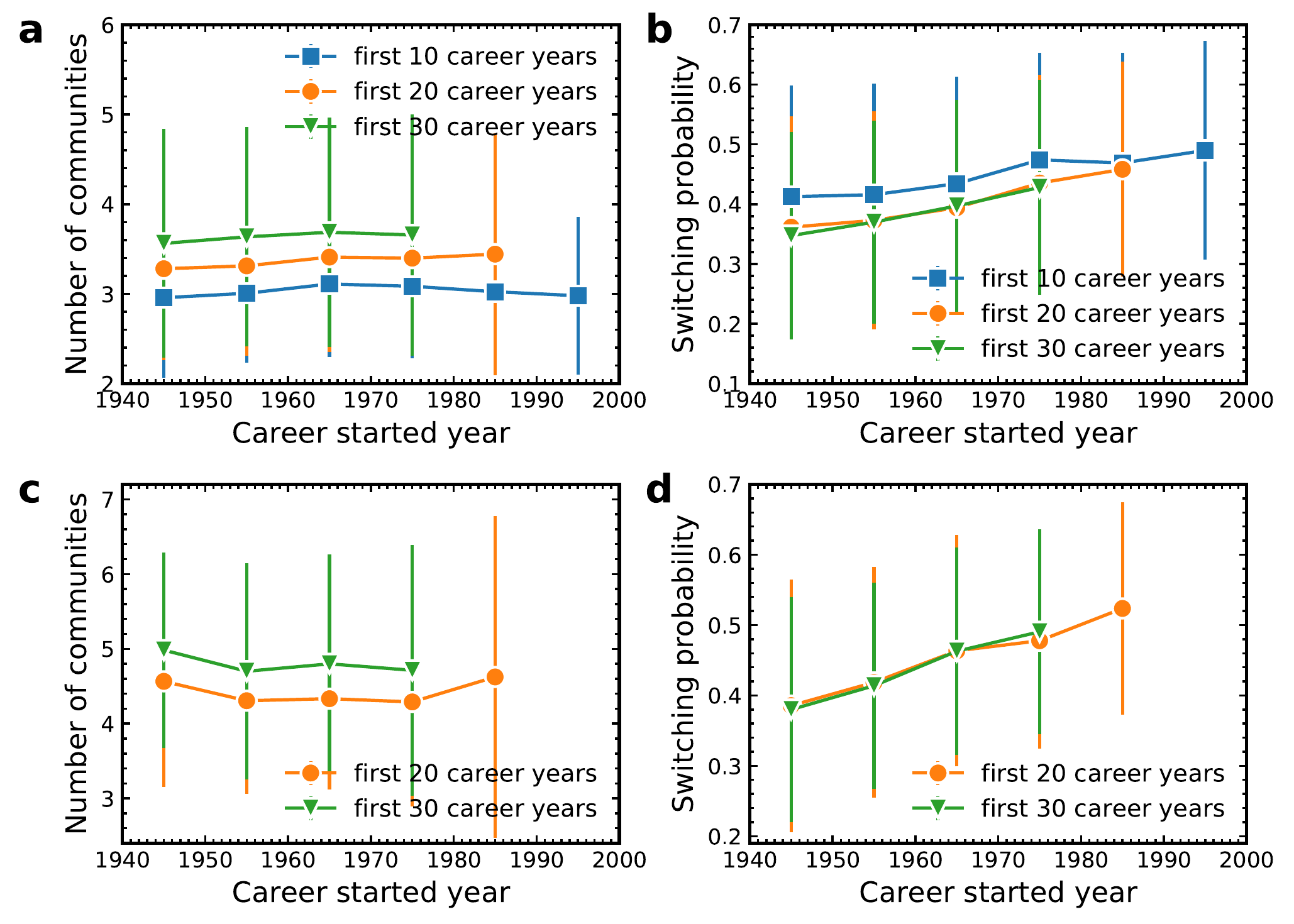}\\
  \textbf{Figure S11.} When we study the evolution of the structural and dynamical properties of CCNs as the development of science, we only consider scientists with at least 30 papers in their first $y$ career years ($y=10, 20, 30$) in order to ensure that the network is sufficiently large to obtain meaningful community detection results. Here, we test two alternative cases where we require the scientists to have at least 20 (see Fig. S11ab) and 50 (See Fig. S11cd) papers in their first $y$ career years. (a)(c) The mean number of communities for scientists who started their career in different years. (b)(d) The average switching probability of scientists who started their career in different years. The results for $y=10$ are not included in (c) and (d) as there are very few scientists who published over 50 papers in their first 10 career years.\label{FigS11}
\end{figure}

\begin{figure}[h!]
  \centering
  \includegraphics[width=13cm]{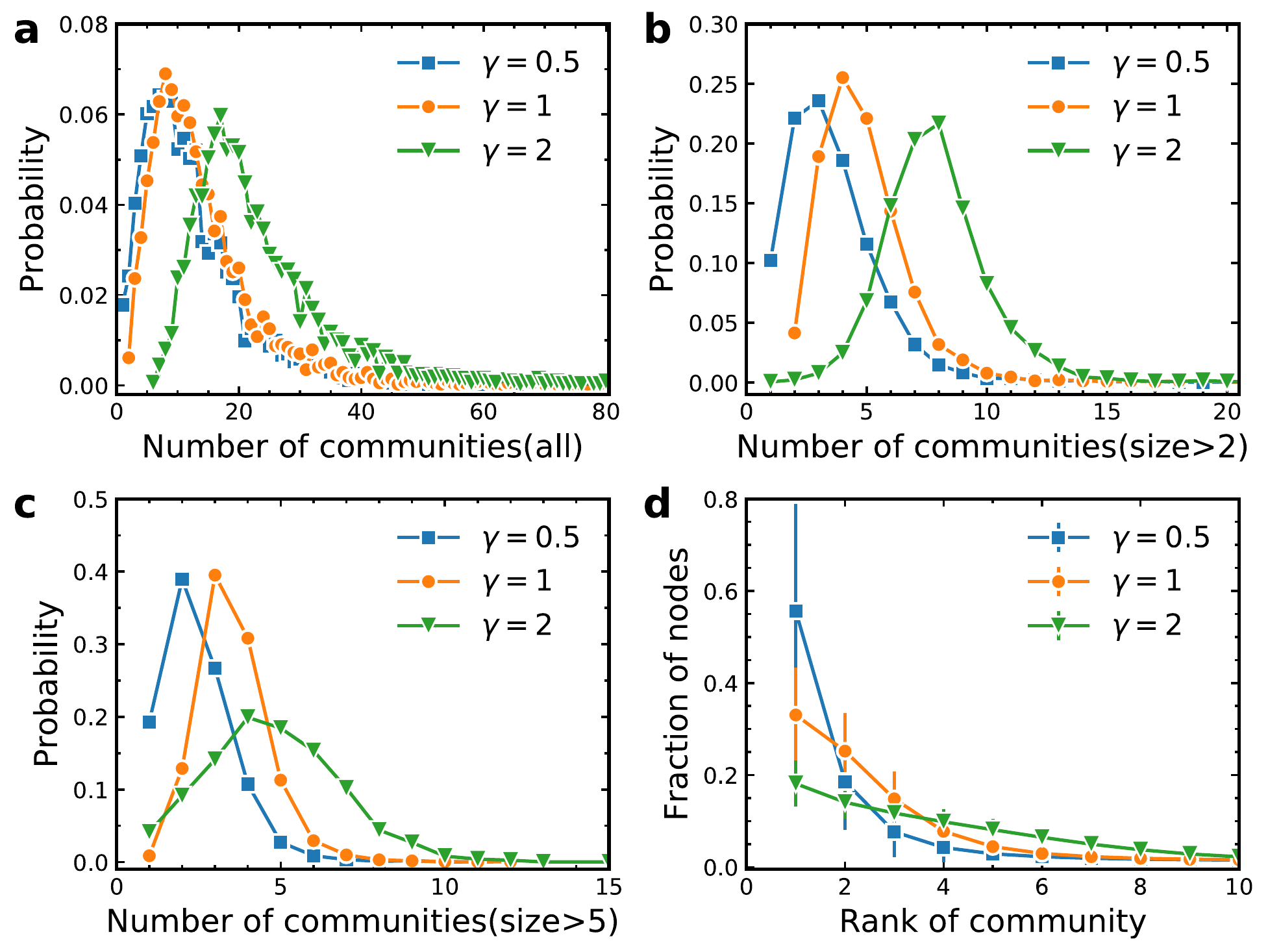}\\
  \textbf{Figure S12.} In the paper, we perform community detection by maximizing the standard modularity function. Here, we test the results with the modified modularity function with a resolution parameter $\gamma$ where a larger $\gamma$ yields more communities. Apart from the standard $\gamma=1$, we consider two typical values of $\gamma$ (i.e. $\gamma=0.5$ and $\gamma=2$) as suggested in ref.~\cite{statistical2006reichardt}. (a) The distributions of the number of communities for all scientists under different resolution parameter $\gamma$. (b) The distribution of the number of communities (size$>$2) for all scientists under different resolution parameter $\gamma$. (c) The distribution of the number of communities (size$>$5) for all scientists under different resolution parameter $\gamma$. Although the number of communities is influenced by the parameter $\gamma$, the distributions after filtering out small clusters (e.g. size$>$2 and size$>$5) are still narrow. (d) Fraction of nodes in different communities under different parameter $\gamma$. \label{FigS12}
\end{figure}

\begin{figure}[h!]
  \centering
  \includegraphics[width=16cm]{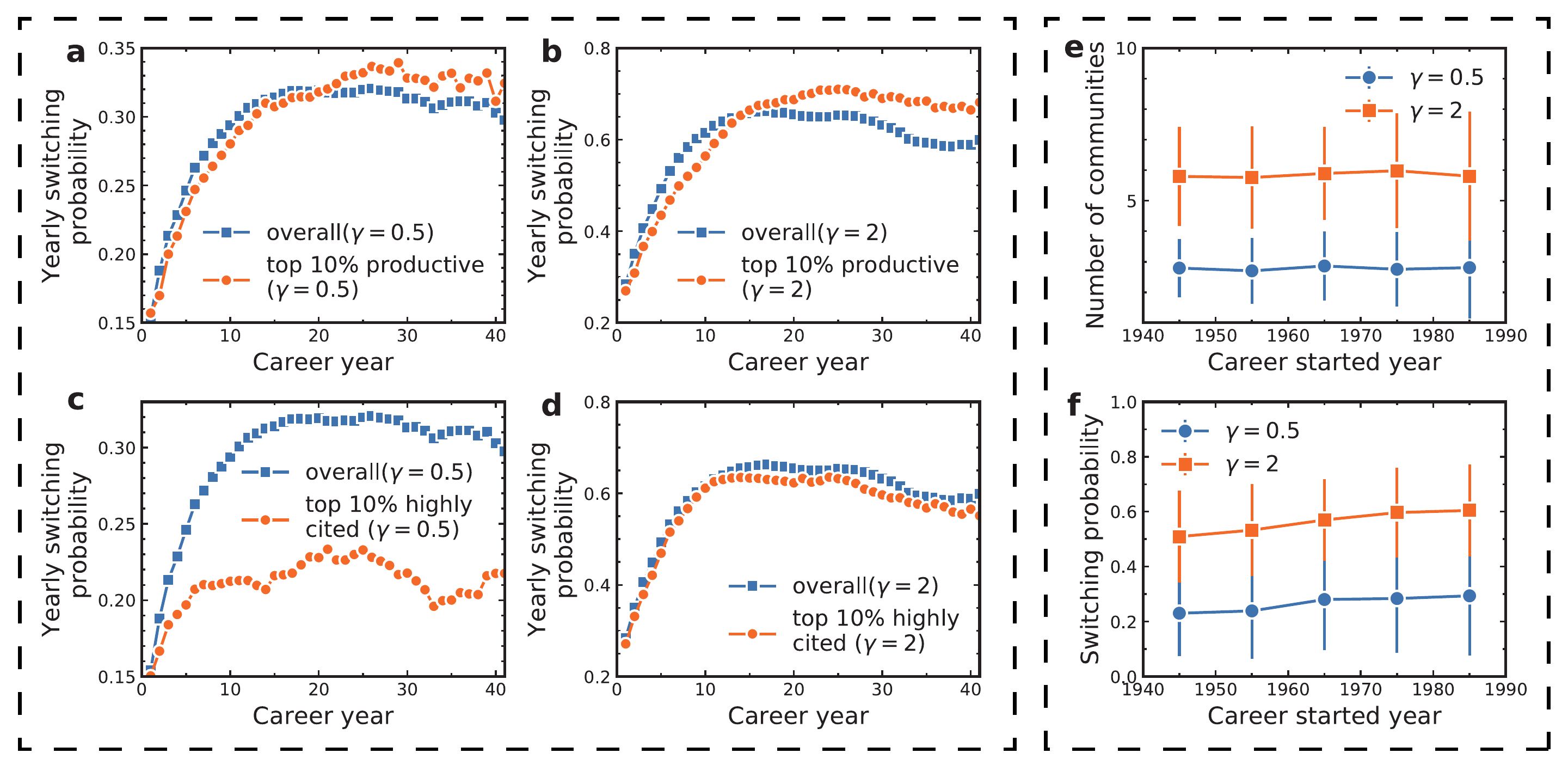}\\
  \textbf{Figure S13.} In the paper, we perform community detection by maximizing the standard modularity function. Here, we test the results with the modified modularity function with a resolution parameter $\gamma$ where a larger $\gamma$ yields small but more communities, and vice versa. (a)(b) Comparison of the overall switching probability with the switching probability of the 10\% most productive scientists in different career years. (c)(d) Comparison of the overall switching probability with the switching probability of the 10\% scientists who has the highest mean citation per paper. $\gamma=0.5$ in (a)(c), while $\gamma=2$ in (b)(d). The large gap in (c) suggests that frequent switching between very dissimilar topics may cause significantly adverse effect on mean citation per paper. (e) The mean number of communities for scientists who started their career in different years (showing both $\gamma=0.5$ and $\gamma=2$ cases). (f) The average switching probability of scientists who started their career in different years (showing both $\gamma=0.5$ and $\gamma=2$ cases). In (e)(f), we only consider scientists' first 20 career years when we compare scientists who started their careers in different years.\label{FigS13}
\end{figure}

\begin{figure}[h!]
  \centering
  \includegraphics[width=12cm]{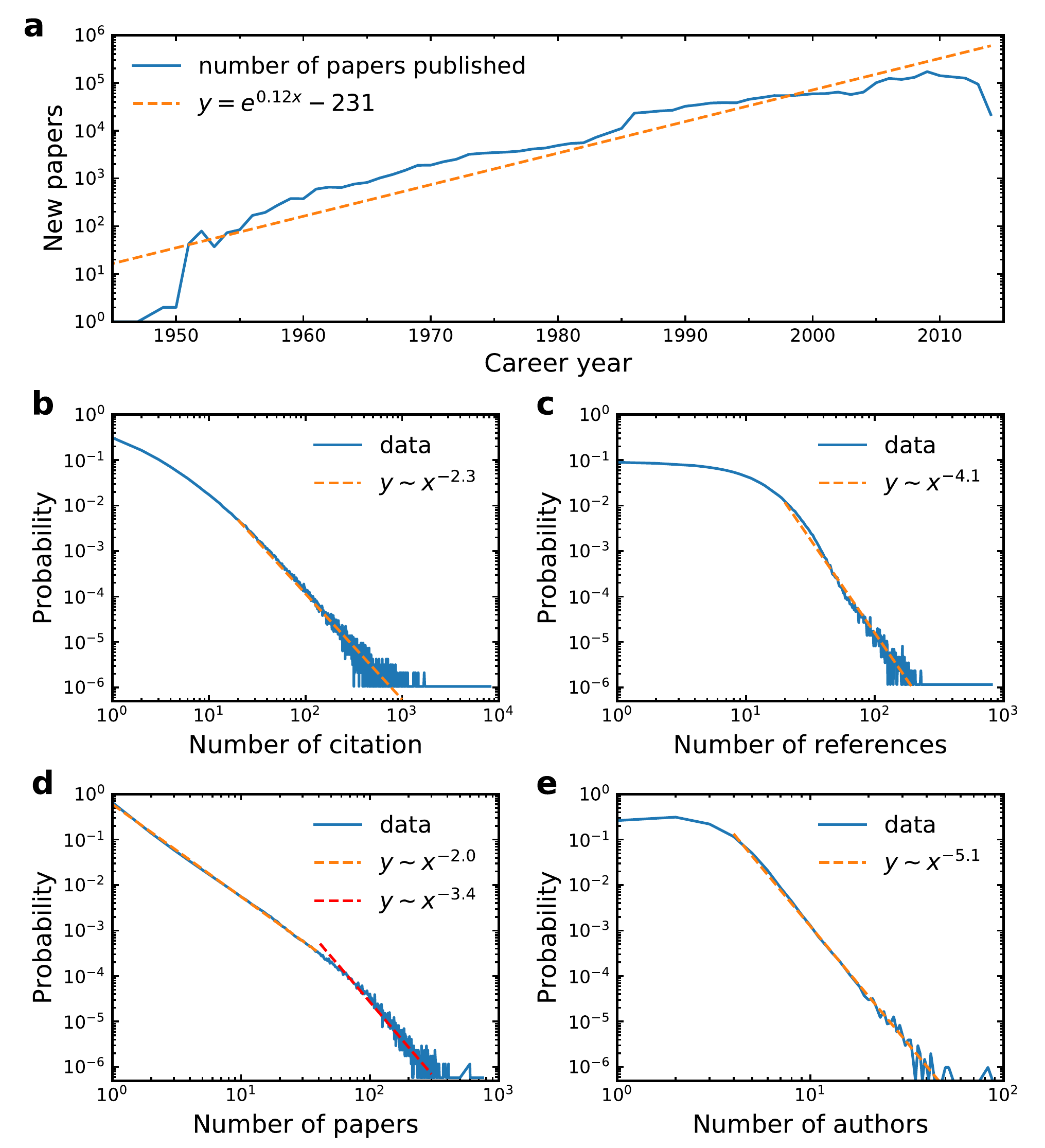}\\
  \textbf{Figure S14.} (a) Exponential fit to the growth of yearly new papers in the computer science data set. (b) Power-law fit to papers' citation distribution. (c) Power-law fit to papers' reference distribution. (d) Double power-law fits to authors' productivity (i.e. number of published papers) distribution for different regimes. (e) Power-law fit to papers' team size (i.e. number of authors) distribution. \label{FigS14}
\end{figure}

\begin{figure}[h!]
  \centering
  \includegraphics[width=16cm]{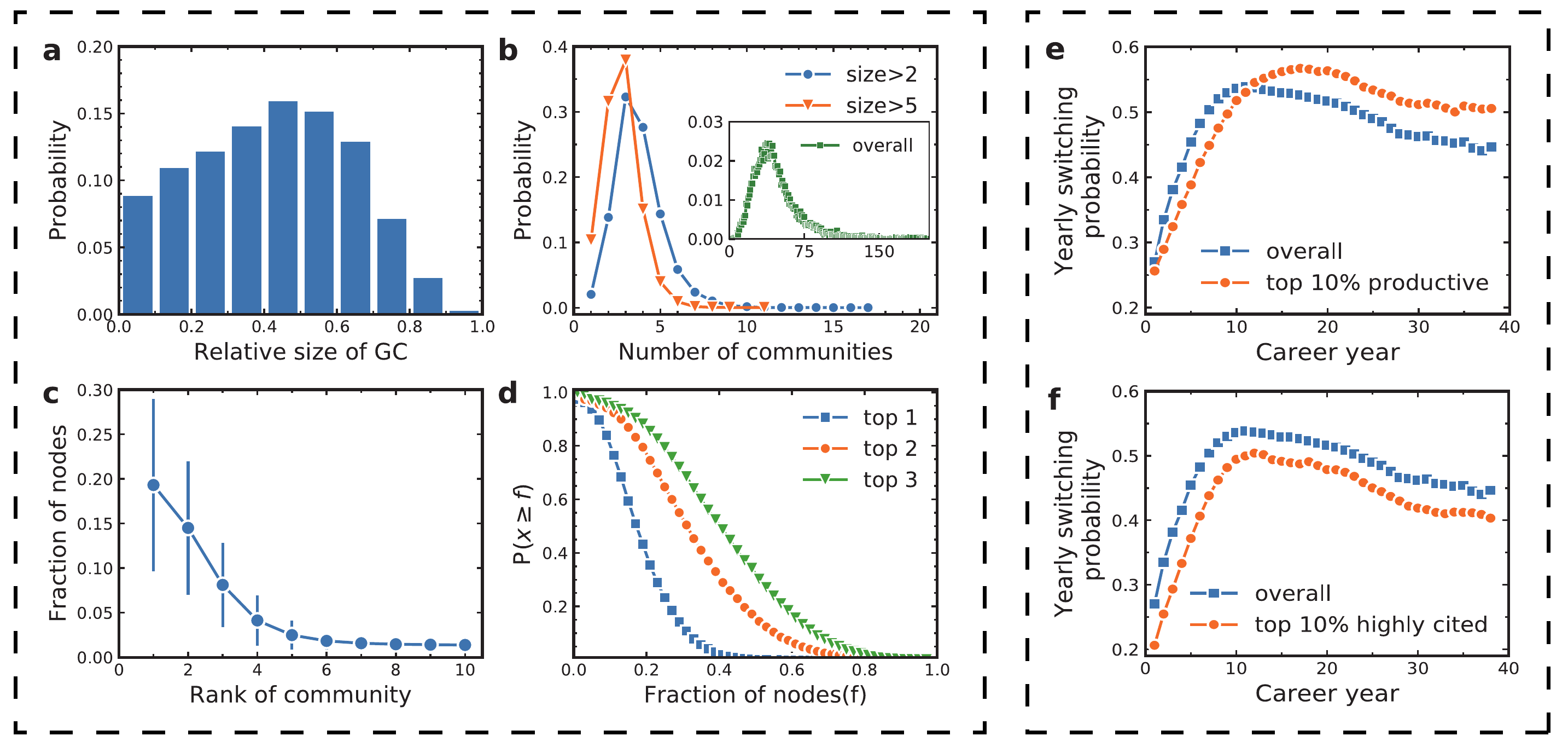}\\
  \textbf{Figure S15.} The structural and switching dynamics analysis of the computer science data. (a) Distribution of the relative size of giant component (GC). The results suggest that the co-citing networks of individual computer scientists are in general less connected than those of physicists. (b) Distribution of the number of communities for all scientists. The inset shows the distribution of the number of communities without filtering any small communities. (c) Fraction of papers in different communities sorted by descending size. (d) Inverse cumulative probability of fraction of nodes in the largest community (legend as ``top 1"), the two largest communities (legend as ``top 2"), and the three largest communities (legend as ``top 3"), respectively. (e) Comparison of the overall switching probability with the switching probability of the 10\% most productive scientists in different career years. (f) Comparison of the overall switching probability with the switching probability of the 10\% scientists who has the highest mean citation per paper. \label{FigS15}
\end{figure}

\begin{figure}[h!]
  \centering
  \includegraphics[width=13cm]{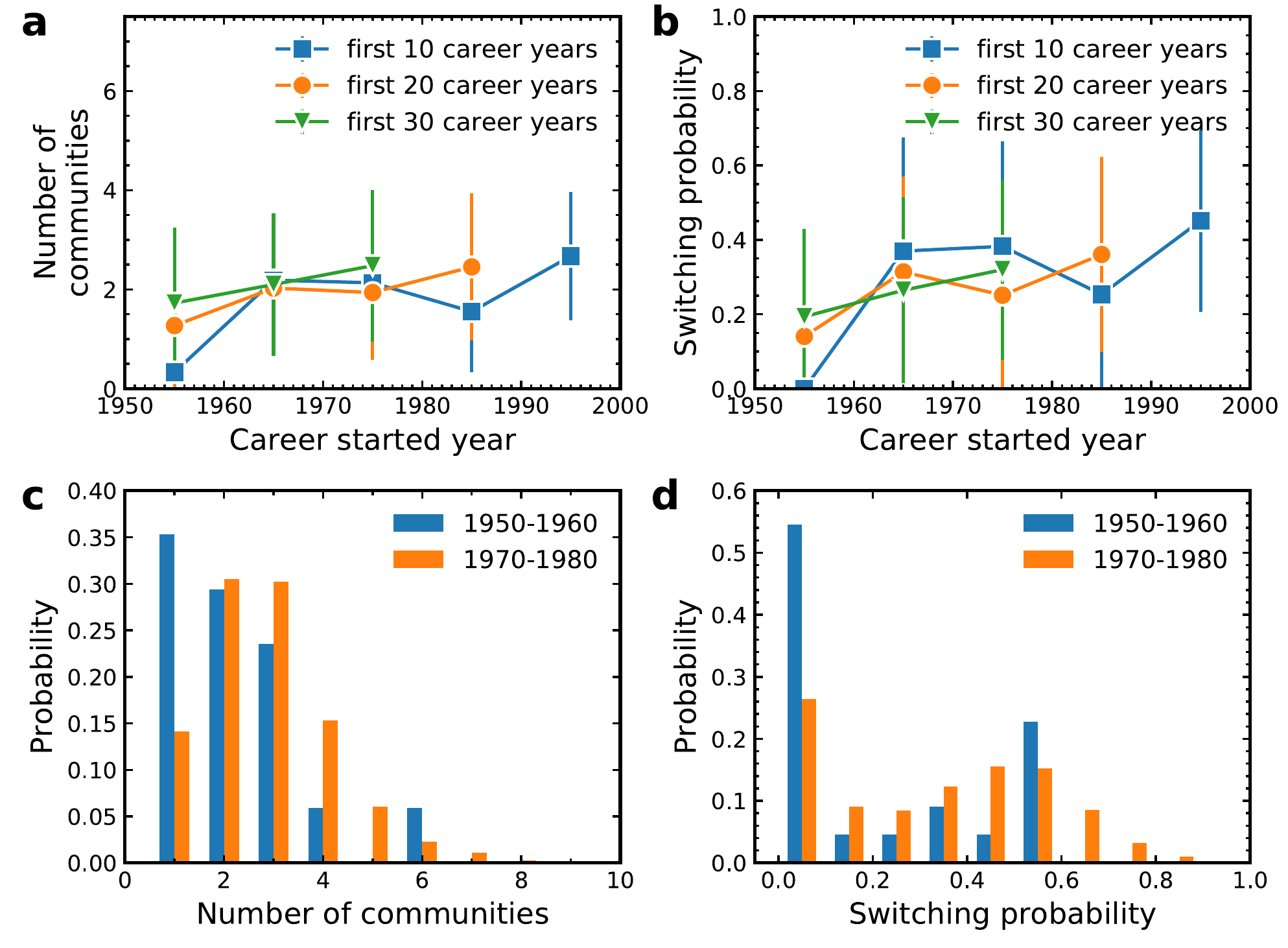}\\
  \textbf{Figure S16.} The evolution of the structural and dynamical properties of CCNs as the development of computer science. Similar to the APS data, we only consider scientists' first $y$ career years and remove (i) all the scientists who not yet reached $y$ years career, and (ii) those who published less than 30 papers in their first $y$ career years. The results for $y=10, 20, 30$ are presented respectively in this figure. (a) The mean number of communities for scientists who started their career in different years. The increasing trend is because the co-citing networks of computer scientists are in general very sparse and thus have many isolated nodes. Papers become more connected in the interdisciplinary era, resulting in the connection of isolated nodes and thus a higher number of clusters with size larger than two. (b) The average switching probability of scientists who started their career in different years. The increasing trend suggests that computer scientists switch between different communities more frequently nowadays. The large fluctuation in $y=10$ and $y=20$ is because there are only a small number of scientists who published over 30 papers in their first 10 or 20 career years. (c) Distributions of the number of communities (for $y=30$) for scientists who started their career between 1950 and 1960, and for those who started their career between 1970 and 1980. The p-value of the Kolmogorov-Smirnov test is $0.098$.(d) Distributions of the switching probability (for $y=30$) of scientists who started their career between 1950 and 1960, and of those who started their career between 1970 and 1980.  The p-value of the Kolmogorov-Smirnov test is $0.034$.\label{FigS16}
\end{figure}

\end{document}